\documentclass[11pt]{article}

\usepackage[final]{acl}

\usepackage{times}
\usepackage{latexsym}

\usepackage[T1]{fontenc}
\usepackage[utf8]{inputenc}

\usepackage{microtype}

\usepackage{inconsolata}

\usepackage{graphicx}

\usepackage{pgfplots}
\pgfplotsset{compat=1.18}

\usepackage{times}
\usepackage{latexsym}

\usepackage{graphicx}
\usepackage{subcaption} % For subfigures
\usepackage{multirow}
\usepackage{enumitem} % For better list control (itemize, enumerate)
\usepackage{booktabs}
\usepackage{amsfonts} % or amssymb
\usepackage{makecell}
\usepackage{float}

\usepackage{algorithm}
\usepackage{algpseudocode} % from algorithmicx
\usepackage{xspace}

\usepackage{amsmath,amssymb,amsthm}

\usepackage{mathtools}
\usepackage{microtype}
\usepackage{bm}
\theoremstyle{definition}

\newtheorem{theorem}{Theorem}
\newtheorem{corollary}{Corollary}
\newtheorem{assumption}{Assumption}
\newtheorem{observation}{Observation}

\newcommand{\F}{\mathcal{F}}
\newcommand{\R}{\mathcal{R}}
\newcommand{\mem}{\text{mem}}
\newcommand{\non}{\text{non}}
\DeclareMathOperator{\E}{\mathbb{E}}
\DeclareMathOperator{\Prb}{\mathbb{P}}

\title{What Hard Tokens Reveal: Exploiting Low-confidence Tokens for Membership Inference Attacks against Large Language Models}

\author{
  Md Tasnim Jawad\textsuperscript{1},
  Mingyan Xiao\textsuperscript{2},
  Yanzhao Wu\textsuperscript{1} \\
  \textsuperscript{1}Florida International University, Miami, FL, USA \\
  \textsuperscript{2}California State Polytechnic University, Pomona, CA, USA \\
  \texttt{mjawa009@fiu.edu},
  \texttt{mxiao@cpp.edu},
  \texttt{yawu@fiu.edu}
}

\newcommand{\prjctname}{HT-MIA}

\begin{document}
\maketitle
\begin{abstract}
With the widespread adoption of Large Language Models (LLMs) and increasingly stringent privacy regulations, protecting data privacy in LLMs has become essential, especially for privacy-sensitive applications. 
Membership Inference Attacks (MIAs) attempt to determine whether a specific data sample was included in the model training/fine-tuning dataset, posing serious privacy risks. 
However, most existing MIA techniques rely on aggregated sequence-level statistics, which conflate generalization with memorization, leading to low attack effectiveness. 
To address this limitation, we propose a novel membership inference approach that captures the token-level probabilities for low-confidence (hard) tokens, where membership signals are more pronounced. 
By comparing token-level probability improvements at hard tokens between a fine-tuned target model and a pre-trained reference model, \prjctname{} isolates strong and robust membership signals that are obscured by prior MIA approaches. 
Extensive experiments on both domain-specific medical datasets and general-purpose benchmarks demonstrate that \prjctname{} consistently outperforms seven established MIA baselines on three of four datasets.
We further investigate differentially private training as an effective defense mechanism against MIAs in LLMs. Overall, our \prjctname{} framework establishes hard-token based analysis as an effective approach for advancing membership inference attacks for LLMs.
\end{abstract}

\section{Introduction}
\label{tab:intro}
In recent years, Large Language Models (LLMs) have become foundational tools to both industry and research. 
These models are frequently fine-tuned for various downstream tasks, including clinical decision support~\cite{wu2023pmc,wang2023clinicalgpt,luo2022biogpt,singhal2023large,paiola2024adapting}, scientific research~\cite{beltagy2019scibert,taylor2022galactica,shi2025deep,rangaraj2026retrieval}, legal document analysis~\cite{chalkidis2020legal,niklaus2024flawn,tewari2024legalpro}, education and intelligent tutoring~\cite{ross2025supervised,scarlatos2025training,hosen2025lora,wu2026computer}, and cybersecurity~\cite{jin2024crimson,ZipZap,das2025security,das2025promptextractionattacksdefenses, sanchez2025semantic,das2026multiturnjailbreakingattackmultimodal}. Many of these settings involve privacy-sensitive or regulated data, making privacy protection a central concern in LLMs. 
A key privacy threat in this context is Membership Inference Attacks (MIAs), which aim to determine whether a specific data sample was included in a model's training or fine-tuning dataset~\cite{shokri2017membership, truex2019demystifying, hu2022membership}. A successful MIA can reveal sensitive information about individuals, institutions, or proprietary datasets, raising serious concerns about regulatory compliance and data privacy. Prior studies have demonstrated the privacy risks posed by MIAs across multiple domains~\cite{truex2019demystifying,zhang2022membership,zhang2023efficient,zhong2024interaction}. Despite these risks, executing effective MIAs against modern LLMs remains challenging. Strong generalization during LLM pre-training and fine-tuning often causes member and non-member samples to exhibit highly similar aggregated prediction behaviors, weakening traditional membership signals~\cite{duan2024membership}. As a result, most existing MIA studies that rely on sequence-level aggregated prediction statistics~\cite{meeus2024did} frequently report low attack success against well-generalized LLMs~\cite{duan2024membership}.

To address this issue, we propose a novel approach that focuses on low-confidence (\textbf{H}ard) \textbf{T}okens for conducting \textbf{M}embership \textbf{I}nference \textbf{A}ttacks (HT-MIA). By comparing token-level probability improvements between a target model and a reference model, our approach isolates strong and robust membership signals that are obscured by prior MIA methods. We benchmark HT-MIA against seven established MIA methods and demonstrate strong performance improvements across four datasets spanning both medical and general domains. In addition, we conduct an ablation study to validate the effectiveness of our design choices, analyze why traditional MIA approaches tend to fail, and investigate potential defense strategies against membership inference risks for LLMs. The source code and experimental configurations are available at \url{https://github.com/006jawad/ht-mia-source}.

\begin{figure*}[!ht]
  \centering
  \includegraphics[width=\linewidth]{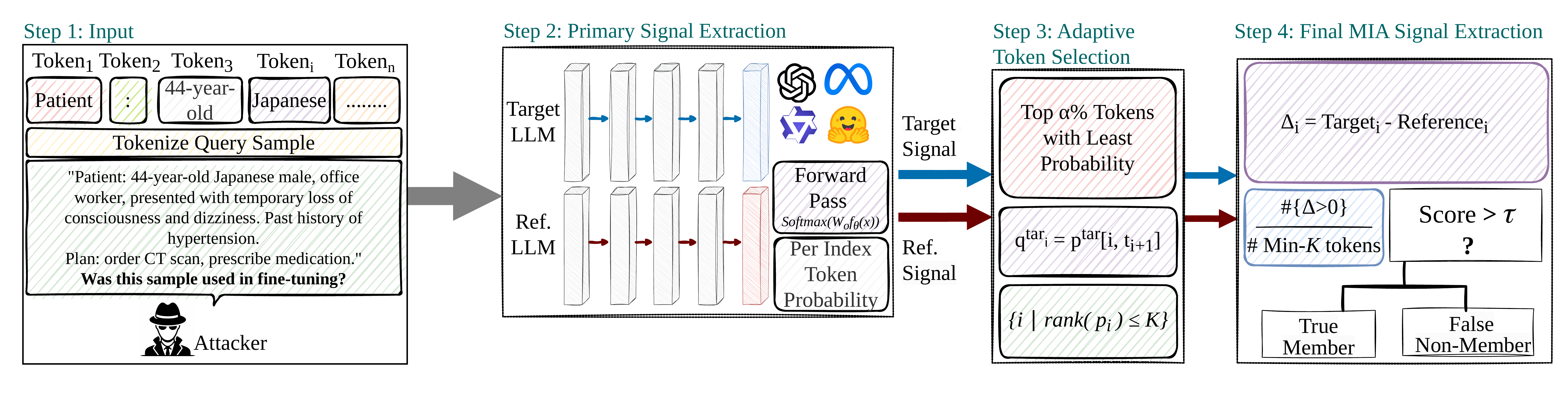}
  \caption{Overall Workflow of \prjctname{}.}
  \label{fig:workflow}
\end{figure*}

\section{Problem Statement}
In this work, we primarily focus on MIAs in the LLM fine-tuning settings. Unlike large-scale LLM pre-training, where datasets are massive, heterogeneous, and often opaque~\cite{achiam2023gpt}, fine-tuning typically adapts a pre-trained LLM to a smaller, task-specific dataset that may contain sensitive or regulated information (e.g., medical data). This setting introduces a more realistic and practically relevant threat model for evaluating MIAs, as membership labels can be clearly defined based on the separation between member and non-member samples in the fine-tuning data split. We note that the proposed \prjctname{} framework is not inherently restricted to fine-tuning and is also applicable to pre-training settings.

\paragraph{Target Model:} The attack target models are obtained by fine-tuning a pre-trained LLM on domain-specific datasets~\cite{jin2023rethinking, jeong2024fine, anisuzzaman2025fine}. These LLMs predict the next token in a sequence conditioned on the preceding context. Given a sample sequence of tokens $x = [x_1, x_2, \dots, x_n]$, the target model assigns a probability distribution over the sequence by factorizing it as  
% \[
% P_\Theta(x) = \prod_{i=1}^{n} P_\Theta(x_i \mid x_{<i}),
% \]  
\begin{equation}
P_\theta(x) = \prod_{i=1}^{n} P_\theta(x_i \mid x_{<i}),
\label{eq:factorization}
\end{equation}
where $P_\theta(x_i \mid x_{<i})$ denotes the prediction probability assigned to token $x_i$ given the prefix $x_{<i}$~\cite{song2024not} and $\theta$ represents the model parameters.
Fine-tuning is performed by minimizing the negative log-likelihood loss:  
% \[
% \mathcal{L}(\Theta) = - \sum_{i=1}^{n} \log P_\Theta(x_i \mid x_{<i}).
% \]  
\begin{equation}
\mathcal{L}(\theta) = - \sum_{i=1}^{n} \log P_\theta(x_i \mid x_{<i}).
\label{eq:nll}
\end{equation}
The ground-truth tokens are provided at each step~\cite{malladi2023fine,chang2024context}. We regard the corresponding pre-trained LLM, prior to fine-tuning, as the reference (Ref.) model $\mathcal{R}$.

\paragraph{Threat Model:} We consider an adversary who aims to infer whether a sample sequence $x$ was included in the training split ($D_{\text{mem}}$) of the fine-tuning dataset $D$ for the target model $\mathcal{F}$. 
The dataset $D$ is partitioned, such that $D = D_{\text{mem}} \cup D_{\text{non-mem}}$ and $D_{\text{mem}} \cap D_{\text{non-mem}} = \varnothing$, where $D_{\text{mem}}$ and $D_{\text{non-mem}}$ represent the member (training) and non-member partitions, respectively. 

The adversaries are assumed to have query access to $\mathcal{F}$, allowing them to obtain model outputs such as probabilities, logits, or generated responses for arbitrary inputs. To conduct membership inference attacks, the adversary extracts a \emph{membership signal} $S(x; \mathcal{F})$ from model outputs. This signal can take different forms, such as the confidence assigned to $x$, the loss value, or a calibrated score derived from comparing multiple models~\cite{he2024difficulty}. The adversary applies a decision rule as 
% \vspace{0.5em}
\begin{equation}
g(x; \mathcal{F}) =
\begin{cases}
\text{member}, & \text{if } S(x; \mathcal{F}) \geq \tau, \\[6pt]
\text{non-member}, & \text{otherwise}.
\end{cases}
\label{eq:decision}
\end{equation} 
% \vspace{0.5em}
Here, $\tau$ is a threshold chosen by the adversary. A prediction of \emph{member} indicates that $x$ is inferred to belong to the training set $D_{\text{mem}}$ ($x \in D_{\text{mem}}$), while \emph{non-member} indicates the opposite~\cite{mireshghallah2022quantifying}. Unlike the free-running mode of generation, an LLM can expose MIA signals under \textit{teacher-forced next-token prediction}~\cite{chang2024context}. In this setting, the model receives the ground-truth sequence prefix $r_{<i}$ as input and is supervised to predict the next ground-truth token $r_i$. This setup allows direct measurement of token-level signals over the sample sequence $x$~\cite{bachmann2024pitfalls}. \prjctname{} therefore operates in a gray-box (open-weight) setting rather than under strict black-box API access: it requires teacher-forced token-level probabilities from the target model and a reference model whose tokenizer is identical to, or exactly alignable with, that of the target.

\newcommand{\algorithmicinput}{\textbf{Input:}}
\newcommand{\algorithmicoutput}{\textbf{Output:}}
\newcommand{\Input}{\Statex \algorithmicinput~}
\newcommand{\Output}{\Statex \algorithmicoutput~}
\algrenewcommand{\algorithmiccomment}[1]{\hfill$\triangleright$~#1}

\begin{algorithm}[!ht]
  \caption{\prjctname{} for a Single Text Input}
  \label{alg:score}
  \begin{algorithmic}[1]
  \footnotesize
    \Input Text $x$; fine-tuned model $\mathcal{F}$; reference model $\mathcal{R}$; 
           shared tokenizer $T$; integers $\texttt{min\_k}$, $\texttt{max\_k}$; 
           limit $\texttt{max\_length}$; proportion $\alpha \in (0,1]$.
    \Output \prjctname{} Score $s \in [0,1]$.
    
    \State $t \gets T.\textsc{encode}(x, \texttt{max\_length})$

    \State $\ell^{\mathcal{F}} \gets \mathcal{F}(t).\textsc{logits}$;\quad $\ell^{\mathcal{R}} \gets \mathcal{R}(t).\textsc{logits}$
    \State $p^{\mathcal{F}} \gets \textsc{Softmax}(\ell^{\mathcal{F}})$;\quad $p^{\mathcal{R}} \gets \textsc{Softmax}(\ell^{\mathcal{R}})$
    
    \For{$i = 1$ \textbf{to} $|t| - 1$}
      \State $y_i \gets t_{i+1}$
      \State $q^{\mathcal{F}}_i \gets p^{\mathcal{F}}[i, y_i]$;\quad $q^{\mathcal{R}}_i \gets p^{\mathcal{R}}[i, y_i]$
    \EndFor \Comment{Per-token probability extraction}
    
    \State $L \gets |t| - 1$
    \State $K \gets \min\!\big(\texttt{max\_k},\ \max(\texttt{min\_k},\ \lfloor \alpha L \rfloor),\ L\big)$
    \State $K \gets \min(K, L)$
    \State \textbf{if} $K = 0$ \textbf{then return} $0.5$  \Comment{Adaptive token selection}
    
    \State $I_{\min} \gets \textsc{ArgsortAscending}(q^{\mathcal{F}})\,[1{:}K]$
    \State $\Delta \gets \big\{ q^{\mathcal{F}}_i - q^{\mathcal{R}}_i \;\big|\; i \in I_{\min} \big\}$
    
    \State $s \gets \dfrac{1}{K} \displaystyle\sum_{i \in I_{\min}} \mathbf{1}\{\Delta_i > 0\}$ \Comment{Membership inference score}
    
    \State \Return $s$
  \end{algorithmic}
\end{algorithm}

\section{\prjctname{} Overview}
\label{tab:framework-overview}
% \vspace{-0.5em}

We present the overall workflow of our proposed \prjctname{} framework in Figure~\ref{fig:workflow}. 
Our method builds on the idea of extracting token-level probabilities for low-confidence (hard) tokens within a given sequence sample. Specifically, we perform a \emph{teacher-forced forward pass} for both the fine-tuned target model $\mathcal{F}$ and the reference model $\mathcal{R}$. The full algorithm is presented in Algorithm~\ref{alg:score}.

The key intuition is that non-member samples behave similarly for both target ($\mathcal{F}$) and reference ($\mathcal{R}$) models, while the fine-tuned model $\mathcal{F}$ assigns systematically higher probabilities to member samples than the reference model $\mathcal{R}$, especially at hard-to-predict tokens with low confidence. Concretely, for a given text prompt $x$:
\begin{itemize}[leftmargin=*, itemsep=2pt, parsep=0pt]
    \item If $x$ is not a member of the training/fine-tuning dataset (\emph{non-member}), $\mathcal{F}$ and $\mathcal{R}$ assign similar probabilities.
    \item If $x$ is a \emph{member} of the training/fine-tuning dataset, $\mathcal{F}$ assigns \emph{higher} probabilities than $\mathcal{R}$, particularly at inherently hard-to-predict tokens.
\end{itemize} 
% Empirically, we observe that tokens with the lowest predicted probability under $\mathcal{F}$ provide the most informative membership signals. 
Improvements at these hard tokens' positions are more indicative of memorization effects introduced during LLM fine-tuning, rather than generalization. In contrast, differences at high-confidence (easy-to-predict) tokens are typically small and carry limited interpretive value for membership inference.

\subsection{Attack Method}
Given a text sample $x$, we perform the following steps:

\noindent\textbf{\textit{Step 1: Token probability extraction.}}

\noindent Given a text sample $x$, the text is first tokenized into a sequence $t_{1:L}$ with corresponding next tokens $y_{1:L}$. Token-level probabilities are then extracted from both the fine-tuned model $\mathcal{F}$ and reference model $\mathcal{R}$ through teacher-forced forward passes. For each position $i = 1, \dots, L$, the following probabilities are obtained:
\setlength\abovedisplayskip{1pt}
\setlength\belowdisplayskip{1pt}
\begin{align*}
p_{\F,i} &= P_{\F}(y_i \mid t_{\le i}), \\
p_{\R,i} &= P_{\R}(y_i \mid t_{\le i}).
\end{align*}

\noindent\textbf{\textit{Step 2: Adaptive token selection.}}
Next, we identify low-confidence token positions by selecting the indices corresponding to the $K$ smallest token probabilities from the target model. Let $I_{\min}$ denote the indices of the $K$ smallest values in $\{p_{\mathcal{F},i}\}_{i=1}^{L}$, where
\begin{equation}
K = \min(\texttt{max\_k},\ \max(\texttt{min\_k},\ \lfloor \alpha L \rfloor)),
\end{equation}
with $\alpha \in (0,1]$ controlling the proportion of selected tokens, and \texttt{min\_k}, \texttt{max\_k} being user-specified lower and upper bounds. This adaptive selection ensures the method scales appropriately with sequence length while remaining robust across both short and long text inputs.

\noindent\textbf{\textit{Step 3: Probability improvement computation.}}

\noindent For each selected token index $i \in I_{\min}$, the probability difference is computed as:
\begin{equation}
\label{eq:delta}
\Delta_i = p_{\F,i} - p_{\R,i}.
\end{equation}
This difference captures the improvement in probability assigned by the fine-tuned target model relative to the reference model.

\noindent\textbf{\textit{Step 4: Membership score aggregation.}}

\noindent The \textbf{\prjctname{}} score is computed by aggregating these per-token comparisons, measuring the fraction of selected tokens where $\mathcal{F}$ outperforms $\mathcal{R}$:
\begin{equation}
S(x; \mathcal{F}, \mathcal{R}) = \frac{1}{K} \sum_{i \in I_{\min}} \mathbf{1}\{\Delta_i > 0\} \in [0, 1].
\end{equation}
A higher score indicates that the fine-tuned model assigns consistently higher probabilities than the reference model at difficult positions, providing evidence of membership. The final decision rule is given by:
\begin{equation}
g(x; \mathcal{F}) = \text{member} \quad \text{iff} \quad S(x; \mathcal{F}, \mathcal{R}) \ge \tau,
\end{equation}
where $\tau$ is a threshold chosen based on the desired false-positive rate. We provide further analysis in Appendix~\ref{sec:appendix_A}.

\subsection{Defense Method}
Considering the critical risks of membership inference attacks, it is essential to explore defenses that mitigate privacy leakage in \prjctname{}. Here, we investigate Differentially Private (DP) training as a practical defense against MIA. DP training provides formal privacy guarantees by limiting the influence of individual training examples through gradient clipping and noise injection~\cite{abadi2016deep}. A randomized mechanism $\mathcal{M}$ is said to satisfy $(\varepsilon, \delta)$-differential privacy if for any two adjacent datasets $D$ and $D'$ differing by one record, and 
any output set $\mathcal{S}$:
$$P[\mathcal{M}(D) \in \mathcal{S}] \leq e^{\varepsilon} \cdot P[\mathcal{M}(D') \in \mathcal{S}] + \delta,$$
where $\varepsilon$ controls the privacy loss (smaller $\varepsilon$ means stronger privacy) and $\delta$ is the probability of the guarantee failing. For each training sample $x_i$ in a batch, per-example gradients are first computed as:
$$\mathbf{g}_t(x_i) = \nabla_{\theta_t} \mathcal{L}(\theta_t, x_i).$$
These gradients are then clipped to bound their $\ell_2$ norm by a threshold $C$:
$$\bar{\mathbf{g}}_t(x_i) \leftarrow \mathbf{g}_t(x_i) / \max\left(1, \frac{\|\mathbf{g}_t(x_i)\|_2}{C}\right).$$
Finally, calibrated Gaussian noise is added to the aggregated clipped gradients:
$$\tilde{\mathbf{g}}_t \leftarrow \frac{1}{b}\left(\sum_i \bar{\mathbf{g}}_t(x_i) + \mathcal{N}(0, \sigma^2 C^2 \mathbf{I})\right),$$
where $\sigma$ is the noise multiplier, $C$ is the clipping norm, and $b$ is the 
batch size, which are carefully tuned to balance model utility and privacy protection~\cite{abadi2016deep}.

\section{Experimental Analysis}
\label{resMain}

% \section{Experimental Setup}

% \paragraph{Target Model Configurations:} To evaluate our \prjctname{} framework, we experiment with $3$ instances of the target model $\mathcal{F}$, spanning different sizes and architectures. Specifically, we fine-tuned GPT-2~\cite{radford2019language}, Qwen-3-0.6B~\cite{yang2025qwen3}, and LLaMA-3.2-1B~\cite{dubey2024llama} as the target models. For our main experiments, we use the corresponding pre-trained base versions of these models as the reference model. In addition, we evaluate a more practical scenario where the exact pre-trained version of the target model is not available. In this scenarios, we compared the target model with other reference models, including Qwen2.5-0.5B~\cite{yang2025qwen3} and DistilGPT-2~\cite{sanh2019distilbert}. Furthermore, to demonstrate the scalability of \prjctname{} on larger models, we extend our evaluation to LLaMA-3.1-8B and LLaMA-3.1-70B~\cite{dubey2024llama}, showcasing the framework's effectiveness beyond the small-scale target model regime. 

\paragraph{Target Model Configurations:} To evaluate our \prjctname{} framework, we conduct experiments with 5 LLMs, spanning different sizes and architectures. Specifically, we fine-tuned GPT-2~\cite{radford2019language}, Qwen-3-0.6B~\cite{yang2025qwen3}, LLaMA-3.2-1B~\cite{meta2024llama32}, LLaMA-3.1-8B~\cite{dubey2024llama}, and LLaMA-3.1-70B~\cite{dubey2024llama} as the target models ($\mathcal{F}$). For our main experiments, we use the corresponding pre-trained base versions as the reference model ($\mathcal{R}$). In addition, we evaluate a more practical scenario in which the exact pre-trained version of the target model is unavailable. In this scenario, we compare the target model against alternative reference models from the same or related model family, including Qwen2.5-0.5B~\cite{qwen2024qwen25} and DistilGPT-2~\cite{sanh2019distilbert,huggingface2019distilgpt2}.

\paragraph{Datasets:}
We utilized four public datasets to construct membership and non-membership samples. For the medical domain, we employ the Augmented Clinical
Notes dataset~\cite{bonnet2023augmentedclinicalnotes} (referred to as
Clinicalnotes) and the Asclepius Synthetic Clinical Notes
dataset~\cite{kweon2023asclepius} (referred to as Asclepius). For broader
coverage, we fine-tuned LLMs on the Large Movie Review (IMDB)
dataset~\cite{maas2011imdb} and the Wikipedia text
corpus~\cite{wikihf2022}. This combination enables comprehensive evaluations
of membership inference risks across domain-specific and general-purpose
settings. For all datasets, we adopt a 50:50 split: one half is used for model
fine-tuning and labeled as members, while the held-out half is reserved as
non-member data. This yields equally sized member $D_{\text{mem}}$ and
non-member $D_{\text{non-mem}}$ sets. All metrics are computed as the bootstrap mean over 1000 resamples of the
evaluation set. Per-run standard deviations are reported separately in the seed study (Table~\ref{tab:seed_study}).

\paragraph{Baselines:} We compare \prjctname{} against seven representative MIA baselines spanning major categories, including loss-based (\textbf{Loss}~\cite{yeom2018privacy}), perturbation-based (\textbf{Lowercase}~\cite{carlini2021extracting}, \textbf{PAC}~\cite{ye2024data}, \textbf{SPV-MIA}~\cite{fu2023practical}), token-level (\textbf{Min-K++}~\cite{zhang2024min}), reference-based (\textbf{Ratio}~\cite{carlini2021extracting}), and compression-based (\textbf{Zlib}~\cite{carlini2021extracting}) approaches. The majority of these methods utilize the full input sequence, diluting the MIA signal. Unlike Min-K++, \prjctname{} leverages a reference model to guide the exploitation of informative low-confidence tokens. Compared to Ratio's full-sequence likelihood ratios, \prjctname{} focuses exclusively on low-confidence tokens where membership information is most concentrated. Detailed descriptions for all baseline methods are provided in Appendix~\ref{sec:appendix_B}. We also provide details regarding the evaluation metrics in Appendix~\ref{sec:appendix_C}.

\begin{table*}[!ht]
\centering
\setlength{\tabcolsep}{6pt}

\begin{subtable}[t]{0.32\textwidth}
\centering
\resizebox{\textwidth}{!}{%
\begin{tabular}{l c c c}
    \toprule
    \textbf{Method} & \textbf{LLaMA-3.2-1B} & \textbf{Qwen-3-0.6B} & \textbf{GPT-2} \\
    \midrule
    Loss         & 0.6332 & 0.6322 & 0.5347 \\
    Lowercase    & 0.6137 & 0.6166 & 0.5431 \\
    Min-K++      & 0.6084 & 0.6103 & 0.5324 \\
    PAC          & 0.6156 & 0.6225 & 0.5300 \\
    Ratio        & 0.6749 & 0.6819 & 0.5332 \\
    Zlib         & 0.6380 & 0.6365 & 0.5366 \\
    SPV-MIA & 0.6399 & 0.6491 & 0.5445\\
    \textbf{\prjctname{}} & \textbf{0.7348} & \textbf{0.7312} & \textbf{0.5679} \\
    \bottomrule
\end{tabular}}
\caption{AUC}
\label{tab:auc_across_models_asclepius}
\end{subtable}
\hfill
\begin{subtable}[t]{0.32\textwidth}
\centering
\resizebox{\textwidth}{!}{%
\begin{tabular}{l c c c}
    \toprule
    \textbf{Method} & \textbf{LLaMA-3.2-1B} & \textbf{Qwen-3-0.6B} & \textbf{GPT-2} \\
    \midrule
    Loss         & 0.1970 & 0.1888 & 0.1202 \\
    Lowercase    & 0.1796 & 0.1799 & \textbf{0.1254} \\
    Min-K++      & 0.1623 & 0.1545 & 0.1151 \\
    PAC          & 0.1734 & 0.1775 & 0.1157 \\
    Ratio        & 0.2330 & 0.2205 & 0.1170 \\
    Zlib         & 0.1977 & 0.1905 & 0.1207 \\
    SPV-MIA & 0.2170 &  0.2720 & 0.1310\\
    \textbf{\prjctname{}} & \textbf{0.3707} & \textbf{0.3129} & 0.1109 \\
    \bottomrule
\end{tabular}}
\caption{TPR@FPR=0.1}
\label{tab:tpr01_across_models_asclepius}
\end{subtable}
\hfill
\begin{subtable}[t]{0.32\textwidth}
\centering
\resizebox{\textwidth}{!}{
\begin{tabular}{l c c c}
    \toprule
    \textbf{Method} & \textbf{LLaMA-3.2-1B} & \textbf{Qwen-3-0.6B} & \textbf{GPT-2} \\
    \midrule
    Loss         & 0.0228 & 0.0220 & 0.0131 \\
    Lowercase    & 0.0247 & 0.0236 & 0.0144 \\
    Min-K++      & 0.0182 & 0.0149 & 0.0127 \\
    PAC          & 0.0189 & 0.0194 & 0.0116 \\
    Ratio        & 0.0271 & 0.0250 & 0.0120 \\
    Zlib         & 0.0239 & 0.0225 & 0.0131 \\
    SPV-MIA & 0.0210 & 0.0410 & 0.0090\\
    \textbf{\prjctname{}} & \textbf{0.0817} & \textbf{0.0477} & \textbf{0.0181} \\
    \bottomrule
\end{tabular}}
\caption{TPR@FPR=0.01}
\label{tab:tpr001_across_models_asclepius}
\end{subtable}

\caption{Comparison of AUC, TPR@FPR=0.1, and TPR@FPR=0.01 across LLaMA-3.2-1B, Qwen-3-0.6B, and GPT-2 fine-tuned on the Asclepius dataset.}
\label{tab:allmetrics_asclepius}
\end{table*}

\begin{table*}[!ht]
\centering
\setlength{\tabcolsep}{6pt}
\begin{subtable}[t]{0.32\textwidth}
\centering
\resizebox{\linewidth}{!}{%
\begin{tabular}{l c c c}
    \toprule
    \textbf{Method} & \textbf{LLaMA-3.2-1B} & \textbf{Qwen-3-0.6B} & \textbf{GPT-2} \\
    \midrule
    Loss         & 0.8212 & 0.8100 & 0.6238 \\
    Lowercase    & 0.7511 & 0.7332 & 0.5651 \\
    Min-K++      & 0.8000 & 0.7649 & 0.6336 \\
    PAC          & 0.7982 & 0.8001 & 0.5886 \\
    Ratio        & 0.8642 & 0.8681 & 0.6440 \\
    Zlib         & 0.7998 & 0.7838 & 0.5954 \\
    SPV-MIA      & 0.5266 & 0.8359 & 0.6233 \\
    \textbf{\prjctname{}} & \textbf{0.8659} & \textbf{0.8843} & \textbf{0.6978} \\
    \bottomrule
\end{tabular}}
\caption{AUC}
\label{tab:auc_clinicalnotes}
\end{subtable}
\begin{subtable}[t]{0.32\textwidth}
\centering
\resizebox{\linewidth}{!}{%
\begin{tabular}{l c c c}
    \toprule
    \textbf{Method} & \textbf{LLaMA-3.2-1B} & \textbf{Qwen-3-0.6B} & \textbf{GPT-2} \\
    \midrule
    Loss         & 0.5895 & 0.5139 & 0.1997 \\
    Lowercase    & 0.4369 & 0.3973 & 0.1393 \\
    Min-K++      & 0.5201 & 0.3754 & 0.2027 \\
    PAC          & 0.5113 & 0.4848 & 0.1504 \\
    Ratio        & \textbf{0.7070} & 0.6427 & 0.2075 \\
    Zlib         & 0.5370 & 0.4581 & 0.1693 \\
    SPV-MIA      & 0.1120 & 0.4820 & 0.1730 \\
    \textbf{\prjctname{}} & 0.7057 & \textbf{0.6861} & \textbf{0.3223} \\
    \bottomrule
\end{tabular}}
\caption{TPR@FPR=0.1}
\label{tab:tpr01_clinicalnotes}
\end{subtable}
\begin{subtable}[t]{0.32\textwidth}
\centering
\resizebox{\linewidth}{!}{%
\begin{tabular}{l c c c}
    \toprule
    \textbf{Method} & \textbf{LLaMA-3.2-1B} & \textbf{Qwen-3-0.6B} & \textbf{GPT-2} \\
    \midrule
    Loss         & 0.3148 & 0.2640 & 0.0223 \\
    Lowercase    & 0.1990 & 0.1891 & 0.0170 \\
    Min-K++      & 0.2168 & 0.0536 & 0.0279 \\
    PAC          & 0.2143 & 0.2251 & 0.0135 \\
    Ratio        & 0.3883 & 0.2842 & 0.0201 \\
    Zlib         & 0.2663 & 0.2407 & 0.0174 \\
    SPV-MIA      & 0.0120 & 0.1330 & 0.0150 \\
    \textbf{\prjctname{}} & \textbf{0.4253} & \textbf{0.3839} & \textbf{0.0524} \\
    \bottomrule
\end{tabular}}
\caption{TPR@FPR=0.01}
\label{tab:tpr001_clinicalnotes}
\end{subtable}

\caption{Comparison of AUC, TPR@FPR=0.1, and TPR@FPR=0.01 across LLaMA-3.2-1B, Qwen-3-0.6B, and GPT-2 fine-tuned on the Clinicalnotes dataset.}
\label{tab:allmetrics_clinicalnotes}
\end{table*}

% \begin{table*}[t]
% \centering
% \small
% \caption{Fine-tuning hyperparameters for each model.}
% \label{tab:finetune_config}
% \begin{tabular}{lcccccc}
% \toprule
% \textbf{Parameter} & \textbf{GPT-2} & \textbf{Qwen3-0.6B} & \textbf{LLaMA-3.2-1B} & \textbf{LLaMA-3.1-8B} & \textbf{LLaMA-3.1-70B} \\
% \midrule
% Learning Rate           & $1\times10^{-4}$ & $1\times10^{-4}$ & $1\times10^{-4}$ & $2\times10^{-5}$ & $2\times10^{-5}$ \\
% Batch Size (per device) & 8 & 8 & 8 & 4 & 2 \\
% Gradient Accum Steps    & 1 & 2 & 2 & 2 & 4 \\
% Epochs                  & 1 & 1 & 1 & 1 & 1 \\
% Max Seq Length          & 512 & 512 & 512 & 512 & 512 \\
% Weight Decay            & 0.01 & 0.01 & 0.01 & 0.01 & 0.01 \\
% Max Grad Norm           & 1.0 & 1.0 & 1.0 & 1.0 & 1.0 \\
% LR Scheduler            & cosine & cosine & cosine & cosine & cosine \\
% Warmup                  & 10\% steps & 10\% steps & 10\% steps & 10\% steps & 10\% steps \\
% Precision               & fp32 & bf16 & bf16 & bf16 & bf16 \\
% \bottomrule
% \end{tabular}
% \end{table*}

\subsection{Overall Performance}
In this section, we highlight the collective findings of our main experiments.
Overall, \prjctname{} outperforms the seven baselines across three out of four datasets.

On \textbf{Asclepius}, it achieves the highest AUC on all three models: 73.48\% (LLaMA-3.2-1B), 73.12\% (Qwen-3-0.6B), and 56.79\% (GPT-2), surpassing Ratio and Min-K++ by 5-6\% on the first two models. 
For TPR@FPR=0.1, \prjctname{} substantially improves recall on LLaMA-3.2-1B (37.07\% vs. 23.30\%) and outperforms all methods at TPR@FPR=0.01 across all models.

On \textbf{Clinicalnotes}, \prjctname{} achieves the highest AUC: 86.59\% (LLaMA-3.2-1B), 88.43\% (Qwen-3-0.6B), and 69.78\% (GPT-2). 
It shows consistent gains at TPR@FPR=0.01, with improvements of up to 10\% over the best baselines.

On \textbf{Wikipedia}, \prjctname{} establishes clear superiority with 86.2\% (LLaMA-3.2-1B) and 91.0\% (Qwen-3-0.6B), outperforming Ratio by 4.7\% and 7.3\% respectively.

On \textbf{IMDB}, \prjctname{} remains highly competitive with 92.3\% (LLaMA-3.2-1B) and 94.7\% (Qwen-3-0.6B), trailing only Ratio while surpassing all other methods by a clear margin.

\subsection{Medical Dataset}
\label{sec:med}
\paragraph{Asclepius dataset:} Table~\ref{tab:allmetrics_asclepius} reports the performance of \prjctname{} against baseline attack methods on LLaMA-3.2-1B, Qwen-3-0.6B, and GPT-2 fine-tuned on Asclepius dataset.
Highlighted in Table~\ref{tab:auc_across_models_asclepius}, \prjctname{} achieves the highest AUC across all three target models. On LLaMA-3.2-1B and Qwen-3-0.6B, our method attains 73.48\% and 73.12\%, surpassing the strongest baseline (Ratio) by 5.99\% and 4.93\%. For GPT-2, \prjctname{} achieves 56.79\%, outperforming all baselines by over 2\%. No baseline matches this consistency across architectures, as our probability improvement metric isolates membership signals that single-model and perturbation-based methods miss.

At relaxed (FPR=0.1, Table~\ref{tab:tpr01_across_models_asclepius}) and strict (FPR=0.01, Table~\ref{tab:tpr001_across_models_asclepius}) thresholds, \prjctname{} consistently outperforms most baselines. At FPR=0.1, our method achieves 37.07\% and 31.29\% on LLaMA-3.2-1B and Qwen-3-0.6B respectively, improving over the next-best methods by 59.1\% and 15.0\%. At the stricter FPR=0.01, \prjctname{} reaches 8.17\%, 4.77\%, and 1.81\% on the three models, achieving nearly $4\times$ the TPR of Min-K++ on LLaMA-3.2-1B and 16.3\% improvement on Qwen-3-0.6B. 

Several key patterns emerge from our analysis. \textit{First}, the performance gap between \prjctname{} and baselines widens as the false positive constraint becomes stricter, indicating more confident and well-separated scores. \textit{Second}, larger models exhibit stronger membership signals than GPT-2, likely due to greater memorization capacity. \textit{Third}, while Ratio and Min-K++ consistently rank among top baselines, neither matches our method's ability to isolate membership-specific improvements through adaptive token selection and cross-model comparison.

% \begin{figure*}[!ht]
%   \centering
%   \footnotesize
%   \begin{subfigure}{0.23\textwidth}
%     \includegraphics[width=\linewidth]{llama_wikipedia_latest.pdf}
%     \caption{LLaMA-3.2-1B (Wiki)}
%     \label{fig:llama3_wiki}
%   \end{subfigure}
%   \hspace{1mm}
%   \begin{subfigure}{0.23\textwidth}
%     \includegraphics[width=\linewidth]{qwen_wikipedia_latest.pdf}
%     \caption{Qwen-3-0.6B (Wiki)}
%     \label{fig:qwen_wiki}
%   \end{subfigure}
%   \hspace{1mm}
%   \begin{subfigure}{0.23\textwidth}
%     \includegraphics[width=\linewidth]{llama_imdb_latest.pdf}
%     \caption{LLaMA-3.2-1B (IMDB)}
%     \label{fig:llama3_imdb}
%   \end{subfigure}
%   \hspace{1mm}
%   \begin{subfigure}{0.23\textwidth}
%     \includegraphics[width=\linewidth]{qwen_imdb_latest.pdf}
%     \caption{Qwen-3-0.6B (IMDB)}
%     \label{fig:qwen_imdb}
%   \end{subfigure}

%   \caption{Comparison of multiple MIA methods on target models fine-tuned with Wikipedia and IMDB datasets.}
%   \label{fig:wiki_imdb_mia}
% \end{figure*}

\begin{figure}[!ht]
  \centering
  \footnotesize
  % Wikipedia plots
  \begin{subfigure}{0.4\textwidth}
    \includegraphics[width=\linewidth]{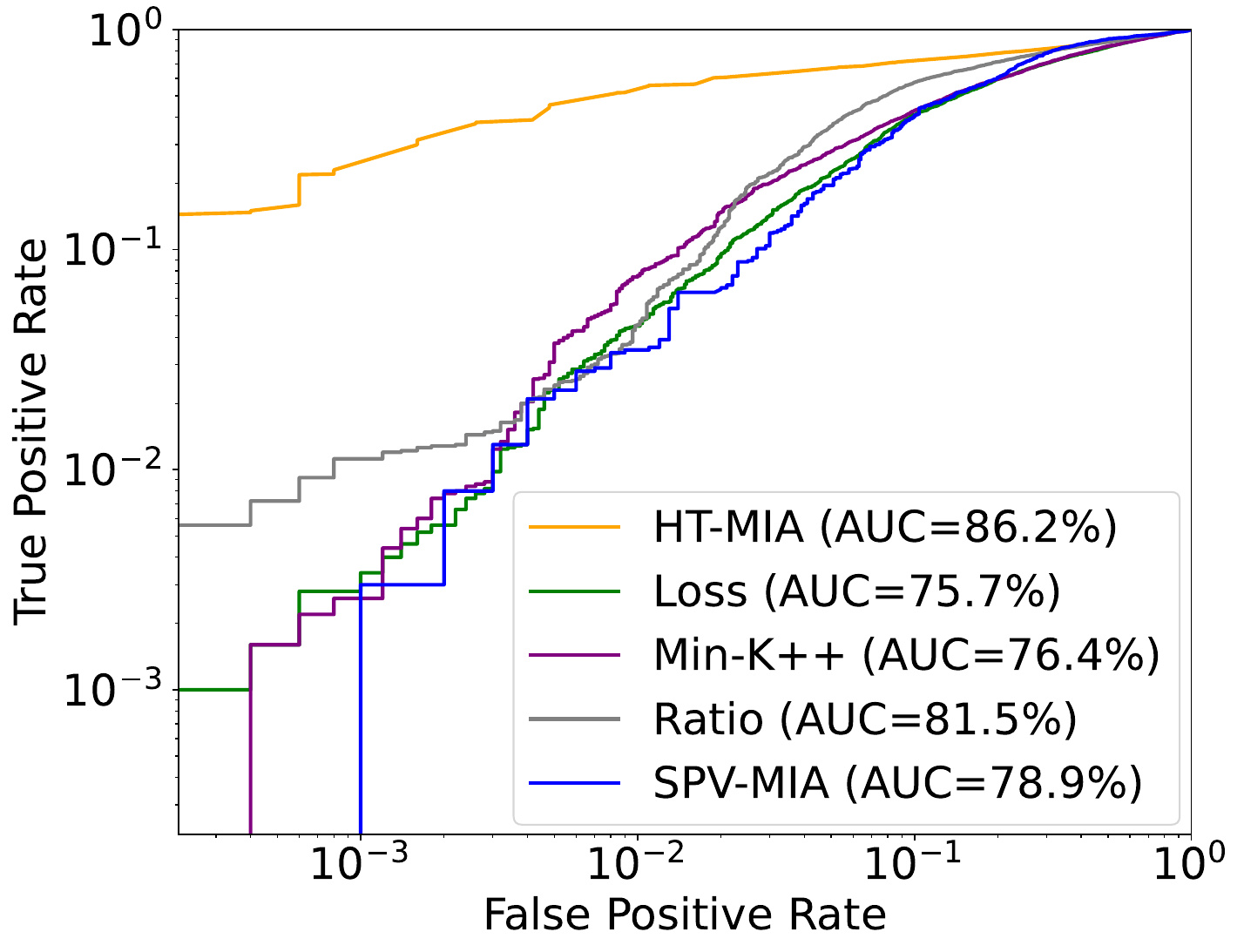}
    \caption{LLaMA-3.2-1B}
    \label{fig:llama3_wiki}
  \end{subfigure}
  \hspace{1mm}
  \begin{subfigure}{0.4\textwidth}
    \includegraphics[width=\linewidth]{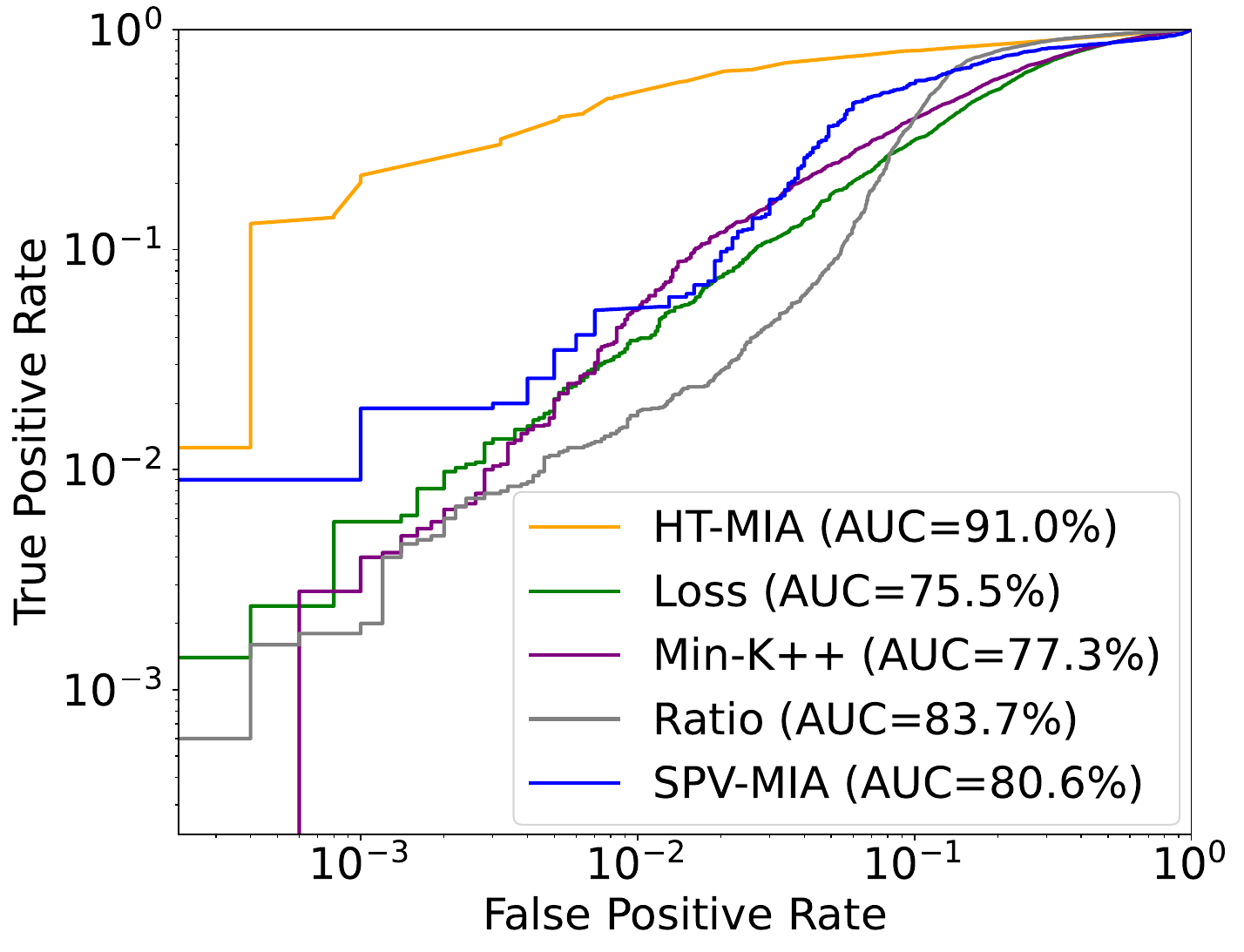}
    \caption{Qwen-3-0.6B}
    \label{fig:qwen_wiki}
  \end{subfigure}
  \caption{Comparison of multiple MIA methods on target models fine-tuned with Wikipedia dataset.}
  \label{fig:wiki_imdb_mia}
\end{figure}

\paragraph{Clinicalnotes dataset:} Table~\ref{tab:allmetrics_clinicalnotes} presents the corresponding results from baseline attack methods on LLaMA-3.2-1B, Qwen-3-0.6B, and GPT-2 fine-tuned on Clinicalnotes dataset.

\begingroup
\looseness=2
As shown in Table~\ref{tab:auc_clinicalnotes}, \prjctname{} achieves the highest AUC across all three target models. On GPT-2, it reaches 69.78\%, exceeding the next-best baseline (Ratio at 64.40\%) by 5.38\%. On LLaMA-3.2-1B and Qwen-3-0.6B, it attains 86.59\% and 88.43\%, surpassing the strongest baseline (Ratio) by 0.17\% and 1.62\%, respectively. The narrower margins on the larger models compared to GPT-2 suggest that the Clinicalnotes dataset produces stronger membership signals overall, making the detection task easier across most methods. Nevertheless, \prjctname{} maintains its advantage by capturing additional discriminative information. The consistently high AUC values on LLaMA-3.2-1B and Qwen-3-0.6B show that fine-tuning on medical text introduces greater distributional shifts than Asclepius. This ranking of \prjctname{} above the baselines is stable across training seeds, as we verify for Qwen-3-0.6B in Appendix~\ref{sec:appendix_seed}.
At both FPR thresholds (Tables~\ref{tab:tpr01_clinicalnotes} and \ref{tab:tpr001_clinicalnotes}), \prjctname{} demonstrates strong performance across all models.
\par
\endgroup

At FPR=0.1, our method leads on Qwen-3-0.6B (68.61\%) and GPT-2 (32.23\%) with relative improvements of 6.7\% and 55.3\% over the best baselines, while Ratio narrowly exceeds \prjctname{} by 0.13\% on LLaMA-3.2-1B (70.70\% vs. 70.57\%). At the stricter FPR=0.01, \prjctname{} attains the highest TPR across all models: 42.53\%, 38.39\%, and 5.24\%.
While Ratio emerges as a strong competitor on larger models with strong membership signals, it falls short of \prjctname{}, particularly at lower FPRs.

\subsection{General Dataset}
\label{sec:gen}
Figure~\ref{fig:wiki_imdb_mia} presents ROC curves comparing \prjctname{} against baseline MIA methods on LLaMA-3.2-1B and Qwen-3-0.6B fine-tuned on the Wikipedia datasets. \prjctname{} achieves the highest AUC across both models: 86.2\% on LLaMA-3.2-1B (Figure~\ref{fig:llama3_wiki}) and 91.0\% on Qwen-3-0.6B (Figure~\ref{fig:qwen_wiki}), outperforming the strongest baseline (Ratio) by 4.7\% and 7.3\%, respectively. The next-best methods trail by 6--9\% on LLaMA-3.2-1B and around 10\% on Qwen-3-0.6B, with traditional metrics like Loss and Zlib showing even weaker performance. The consistently larger margins on Qwen-3-0.6B suggest that our probability improvement metric is particularly effective at exploiting memorization patterns when fine-tuned on encyclopedic text.

% \textbf{Wikipedia dataset: Low-FPR region analysis.}
The ROC curves reveal that \prjctname{}'s performance advantage becomes more pronounced in the low false positive rate region (FPR $<$ 0.01), critical for practical deployment. On both models, our method maintains substantially higher TPR than all baselines when FPR is constrained below 1\%. At FPR $\approx$ 0.001, \prjctname{} achieves approximately 5-10$\times$ higher TPR on LLaMA-3.2-1B and 10-15$\times$ on Qwen-3-0.6B compared to the second-best methods. This separation reflects the more confident, well-separated scores from our cross-model comparison, which is valuable when false positives must be minimized. We provide additional details in Appendix~\ref{sec:appendix_genData}.

\subsection{Comparison with State-of-the-Art MIAs}\label{sec:withSOTA}
We also compare against two recent state-of-the-art MIAs, 
EZ-MIA~\cite{ilic2026powerful} and WBC~\cite{chen2026window}.
\begin{figure}[h]
\centering
\includegraphics[width=0.95\columnwidth]{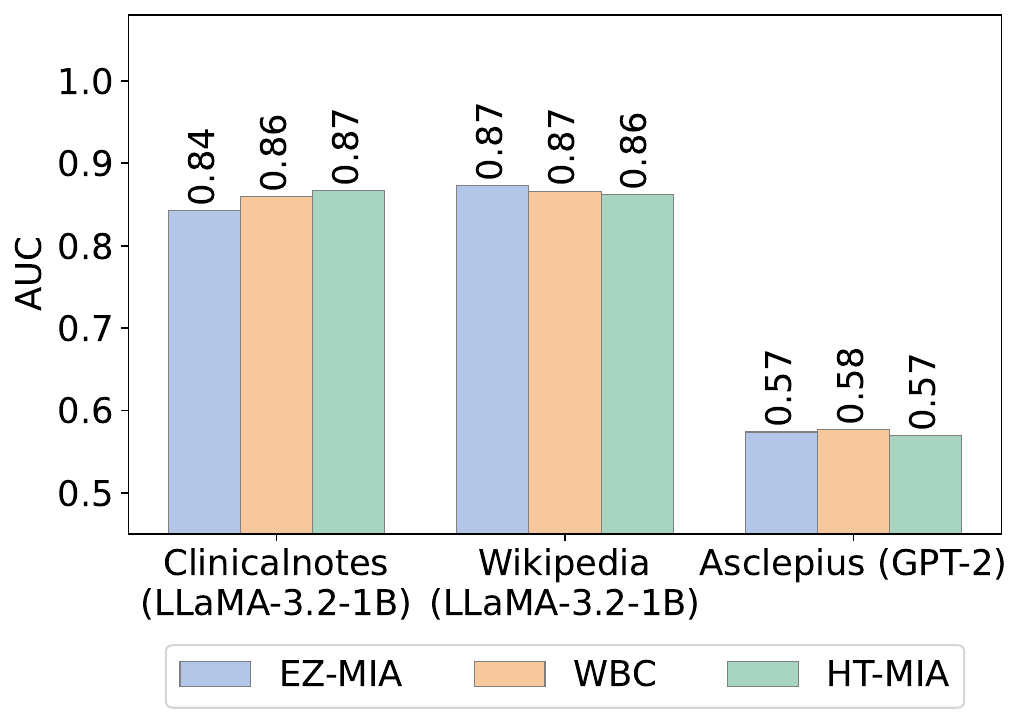}
\caption{AUC comparison of \prjctname{} against EZ-MIA and WBC.}
\label{fig:sota_bar}
\end{figure}

As shown in Figure~\ref{fig:sota_bar}, \prjctname{} is competitive with EZ-MIA and WBC in AUC on several settings and achieves the highest AUC on Clinicalnotes with LLaMA-3.2-1B, where it outperforms EZ-MIA by $2.37$ percentage points. On other datasets, and particularly at low false-positive rates, it often trails these newer methods. Unlike EZ-MIA, which ignores correctly predicted tokens, and WBC, which uses binary windows, \prjctname{} directly scores the least-confident tokens. The main appeal of \prjctname{} is its simplicity and reliance on only two forward passes.
We provide more details in Appendix~\ref{sec:appendix_sota}.

\section{Defense Against \prjctname{}}

% \begin{figure}[t]
%     \centering
%     \includegraphics[width=0.50\textwidth]{combined_radar_chart.pdf}
%     \caption{AUC comparison of MIA methods on GPT-2 and Qwen-3-0.6B fine-tuned with and without DP on the Asclepius dataset.}
%     \label{fig:radar_dpsgd}
% \end{figure}
% \begin{figure}[t]
%     \centering
%     \includegraphics[width=0.50\textwidth]{combined_radar_chart_clinicalnotes_dp.pdf}
%     \caption{AUC comparison of MIA methods on GPT-2 and Qwen-3-0.6B fine-tuned with and without DP on the Clinicalnotes dataset.}
%     \label{fig:radar_dpsgd_clinicalnotes}
% \end{figure}
\begin{figure}[t]
    \centering
    \begin{subfigure}[b]{0.47\textwidth}
        \centering
        \includegraphics[width=\textwidth]{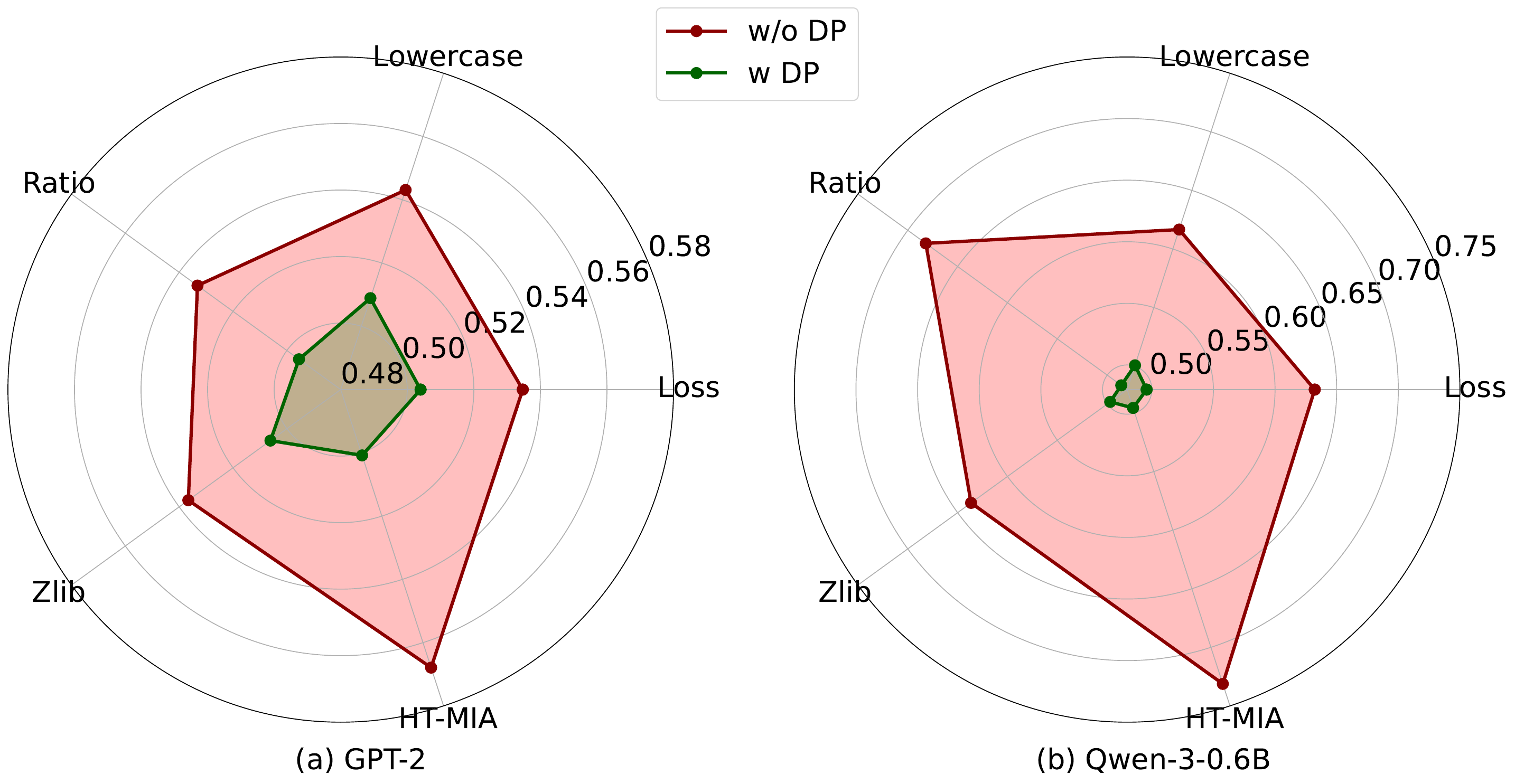}
        \caption{Asclepius}
        \label{fig:radar_dpsgd_asclepius}
    \end{subfigure}
    \hfill
    \begin{subfigure}[b]{0.47\textwidth}
        \centering
        \includegraphics[width=\textwidth]{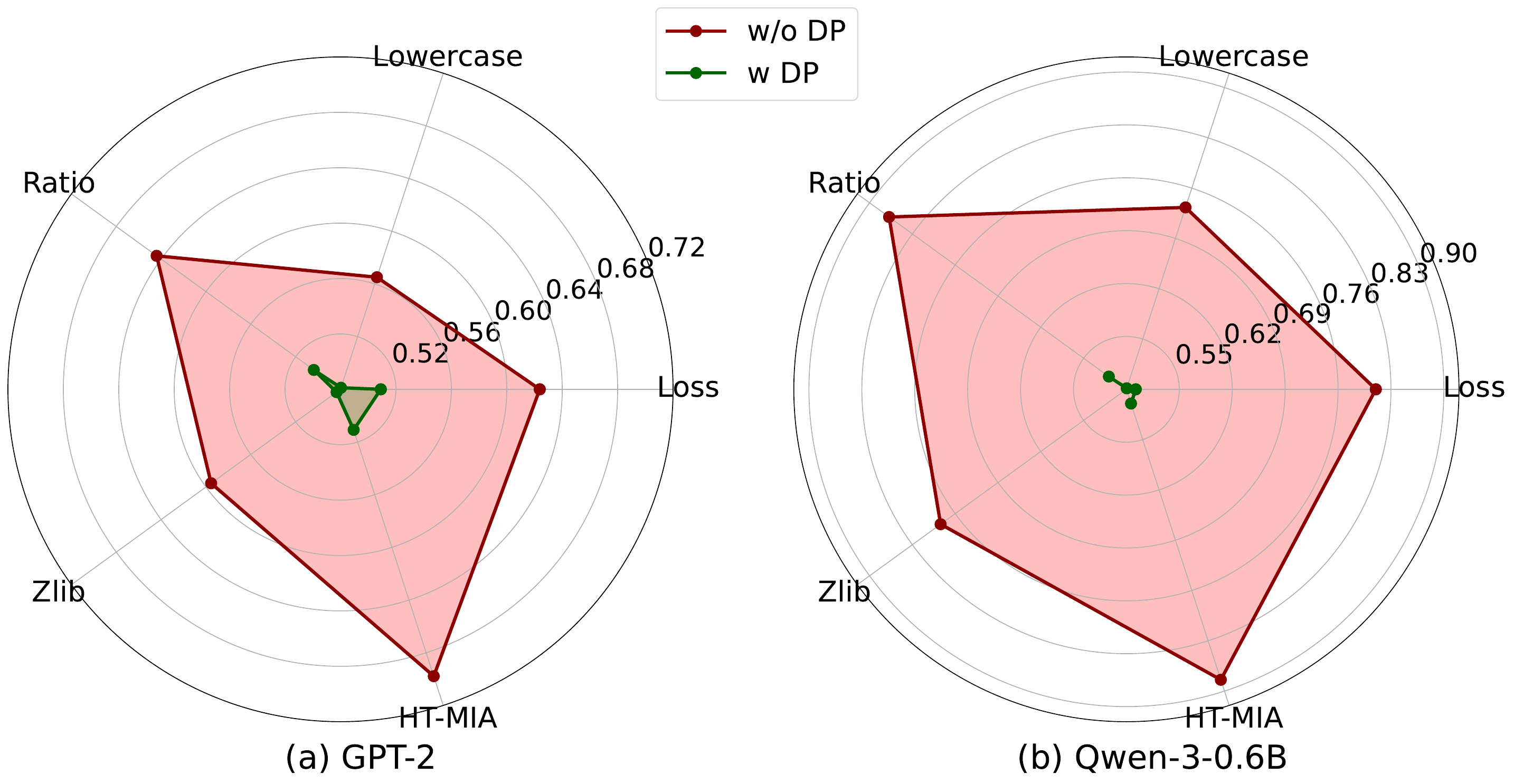}
        \caption{Clinicalnotes}
        \label{fig:radar_dpsgd_clinicalnotes}
    \end{subfigure}
    \caption{AUC comparison of MIA methods on GPT-2 and Qwen-3-0.6B fine-tuned with and without DP on the Asclepius and ClinicalNotes datasets.}
    \label{fig:radar_dpsgd}
\end{figure}

To defend against \prjctname{}, we incorporate DP in our LLM fine-tuning setup to evaluate the MIA defense performance as Figure~\ref{fig:radar_dpsgd} shows. We set $\varepsilon=100$, $\delta=10^{-5}$, and clipping norm $C=100$, with the
noise multiplier auto-calibrated accordingly and all other hyperparameters as in Table~\ref{tab:finetune_config} (batch size $8$, one epoch). We present this setting as a preliminary empirical intervention rather than a formal privacy operating point: at $\varepsilon=100$ with $C=100$ the nominal guarantee is weak, and we do not report a full privacy accounting for it. The accounted analysis is the $C=1$ sweep of Appendix~\ref{sec:appendix_F} across $\varepsilon \in \{1,2,4,8\}$, which we take as the privacy result of record. In all cases, the AUC decreases when DP is applied, reducing the empirical success of the attack. The \prjctname{} AUC falls from 0.5679 to 0.5008 for GPT-2, a drop of 6.71 percentage points (11.8\%), and from 0.7312 to 0.4957 for Qwen-3-0.6B, a drop of 23.55 percentage points (32.2\%). We further validate this on the Clinicalnotes dataset, as shown in Figure~\ref{fig:radar_dpsgd_clinicalnotes}. The \prjctname{} AUC falls from 0.6978 to 0.5108 for GPT-2, a drop of 18.70 percentage points (26.8\%), and from 0.8843 to 0.4998 for Qwen-3-0.6B, a drop of 38.45 percentage points (43.5\%) that removes almost all of the attack's advantage over chance. Together these provide consistent empirical evidence that DP reduces attack success across the two evaluated medical datasets.

This decrease in attack effectiveness indicates that DP reduces the empirical success of membership inference against \prjctname{}. Similar trends are consistently observed across all baseline methods (Loss, Lowercase, Ratio, Zlib), with Qwen-3-0.6B benefiting more from DP than GPT-2. We provide the complete DP-based defense results in Appendix~\ref{sec:appendix_F}.

\section{Related Work}
\label{sec:related}
Existing MIA research on LLMs can be divided into three broad categories: (1) attacks against pre-trained models, (2) attacks against fine-tuned models, and (3) defense mechanisms.
\vspace{-0.5em}
% \textbf{Pretrained models:} 
\paragraph{MIAs against Pre-trained Models:} \cite{duan2024membership, meeus2025sok} benchmark baseline MIA methods on pre-trained LLMs, finding generally poor MIA performance due to diverse pretraining corpora and n-gram overlap. \cite{kaneko2024sampling} proposes SaMIA, a likelihood-independent method using sampling and n-gram overlap. \cite{maini2024llm} introduces dataset-level MIA for detecting entire training datasets. \cite{tao2025token} presents InfoRMIA, a token-level information-theoretic attack that identifies memorized tokens and achieves stronger inference with fine-grained leakage localization. InfoRMIA localises leakage through an information-theoretic criterion with statistical calibration, whereas \prjctname{} uses only the ordering of target-model probabilities to choose positions and the sign of the target--reference gap to score them. \cite{ibanez2024lumia} presents LUMIA, which leverages internal hidden states rather than output likelihoods.
\vspace{-0.5em}
% \textbf{Fine-tuned models:} 
\paragraph{MIAs against Fine-tuned Models:} \cite{huang2025df} proposes DF-MIA, a distribution-free attack tailored for fine-tuned LLMs. \cite{fu2023practical} introduces SPV-MIA, which builds reference datasets by self-prompting. \cite{he2025towards} proposes PETAL, a label-only method using per-token semantic similarity. \cite{song2024not} presents WEL-MIA, which assigns different weights to tokens based on their membership inference difficulty. WEL-MIA thus weights every position by a difficulty-aware criterion with its own reference calibration, while \prjctname{} trades adaptive weighting for a hard deterministic selection of the bottom-$K$ positions and unweighted sign aggregation. Despite these advances, MIA research specifically targeting fine-tuned LLMs remains limited, in part due to the high computational costs associated with fine-tuning LLMs. 
\vspace{-0.5em}
\paragraph{Defenses against MIAs:} A range of defenses have been proposed to mitigate membership inference risks. Existing studies include confidence perturbation (MemGuard)~\cite{jia2019memguard}, purification-based defenses~\cite{yang2020defending}, and training in membership-invariant subspaces (MIST)~\cite{li2024mist}. LLM-specific defenses include pruning-based approaches~\cite{gupta2025pruning}, data transformation methods such as SOFT~\cite{zhang2025soft}, and soft-label training via RM Learning~\cite{zhang2024defending}. Parameter-efficient fine-tuning (PEFT) reduces membership leakage compared with full fine-tuning, while careful hyperparameter selection helps preserve model utility~\cite{ilic2026powerful,jawad2025eftlr}. Differentially Private Stochastic Gradient Descent~\cite{abadi2016deep} provides formal privacy guarantees by bounding the influence of individual training samples and remains a widely adopted defense against MIAs, which we adopt as an effective defense against our proposed \prjctname{}.

\section{Discussion}
\label{sec:discussion}
We evaluate \prjctname{} against seven baseline MIA methods across four popular datasets. As shown in Section~\ref{resMain}, \prjctname{} outperforms these established MIA approaches on three of the four datasets, spanning both domain-specific and general-purpose settings (see Tables~\ref{tab:allmetrics_asclepius},~\ref{tab:allmetrics_clinicalnotes} and Figure~\ref{fig:wiki_imdb_mia}).

Our ablation study (Appendix~\ref{subsec:abltn}) reveals that while token-level probability improvements follow similar trends for members and non-members overall, they diverge sharply at low-confidence tokens (Figures~\ref{fig:token_prob_improve_cln} and~\ref{token_prob_improve_imdb}), justifying our focus on these tokens where membership signals are most concentrated.

MIA effectiveness varies across datasets. On Wikipedia, \prjctname{} achieves substantial improvements (4.7-7.3\% AUC); its articles are the longest and most varied of the four corpora. On Clinicalnotes its advantage over Ratio narrows at relaxed thresholds, particularly for LLaMA-3.2-1B, where all methods reach high AUC and the margins between them are small. Performance on IMDB is slightly lower (-2.7 to -3.6\%); IMDB reviews are short and formulaic relative to the other corpora, and ratio-based methods remain competitive in that setting. Nevertheless, \prjctname{} maintains enhanced performance at strict FPR thresholds across all datasets, demonstrating its value for high-precision membership inference.

\section{Conclusion}
% \vspace{-0.6em}
We make three original contributions in this paper.
\textit{First,} we propose \prjctname{}, a novel token-level membership inference attack that focuses on low-confidence (hard) tokens to capture strong membership signals.
\textit{Second,} we conduct extensive experiments demonstrating that our \prjctname{} approach outperforms seven established MIA baselines on three of four public datasets.
\textit{Third,} through ablation studies, we explain why traditional MIA techniques underperform on fine-tuned LLMs and investigate effective defense strategies against MIAs. Overall, this study highlights critical MIA vulnerabilities in LLMs and demonstrates the importance of fine-grained, token-level analysis for effective membership inference and privacy protection.

\section*{Limitations}
\label{sec:limitations}
In a strict black-box setting, token-level probabilities or other fine-grained MIA-related signals may be inaccessible to the adversary, which limits the direct applicability of \prjctname{} under such access constraints.
Addressing this limitation is important for extending the method to more realistic scenarios where only input samples are available for membership testing.
Another limitation is the reliance on a compatible reference model. \prjctname{} compares token-level probabilities between the target and reference models position by position. This requires an identical or exactly alignable tokenizer. Across model families with mismatched vocabularies, the comparison breaks down and the attack degrades to near-random performance. Even with a matched tokenizer, effectiveness still depends on reference model compatibility.

% \section{Future Work}
In future work, we plan to investigate MIAs under stricter black-box assumptions. We also aim to explore extensions that relax the dependency on closely matched reference models, enabling broader applicability across heterogeneous model families.
Another direction concerns the token-selection rule. Selecting tokens by their largest improvement would reuse the same signal for both selection and scoring, which is why we select by target-model difficulty instead. Selecting by absolute improvement, or scoring positions held out from selection, are alternatives we plan to examine.
Finally, motivated by the strong privacy risks exposed by our findings, we plan to design and evaluate effective MIA defense mechanisms.

% that explicitly mitigate MIA in fine-tuned LLMs.

\section*{Ethical Considerations}
This paper investigates membership inference in LLM to better understand privacy risks associated with LLMs. All experiments use publicly available datasets and open-source models. No human subjects or personally identifiable information are involved.

Although membership inference techniques may raise concerns regarding misuse, our work aims to promote transparency around privacy vulnerabilities and to encourage the development of stronger defenses. We believe that responsible study of such risks is necessary for improving the privacy-preserving deployment of LLM.

\section*{Acknowledgments}
The authors acknowledge partial support from the U.S. National Science Foundation (NSF) grants (IIS-2331908, OAC-2530965, and CREST-2514616), the U.S. Department of Energy (DOE) award (DE-SC0024576), the Defense Advanced Research Projects Agency (DARPA) award (HR00112420004), National Artificial Intelligence Research Resource (NAIRR) Pilot (NAIRR240244 and NAIRR250261); CAHSI-Google Institutional Research Program (IRP); OpenAI; and Amazon Web Services. 
Any opinions, findings, conclusions, or recommendations expressed in this material are those of the authors and do not necessarily reflect the views of the funding agencies and companies mentioned above.
We used OpenAI Codex in a supporting role for implementation, for polishing the language and the presentation of figures, for cross-checking protocols and numerical results, and for literature retrieval. The authors designed the method and experiments, ran all experiments, and made all scientific decisions. The authors also reviewed and tested all AI-assisted code, verified all sources, outputs, and claims, and take full responsibility for the final manuscript and its results.

% Bibliography entries for the entire Anthology, followed by custom entries
%\bibliography{anthology,custom}
% Custom bibliography entries only
\bibliography{custom}

\clearpage
\appendix
In the appendix, we provide additional material and technical details.

% ==================== THEORETICAL ANALYSIS (Alternative without mdframed) ====================
\section{Theoretical Analysis}
\label{sec:appendix_A}
\noindent We present the reasoning behind our design choices and the assumptions under which they are justified. The analysis should be read as principled motivation rather than a proof of optimality: the assumptions below are idealisations of the autoregressive setting, and we state explicitly what each does and does not establish. Our method makes three key design decisions: (1) using binary indicators $\mathbf{1}\{\Delta_i > 0\}$ rather than raw probability differences, which provides robustness to outliers and yields interpretable scores in $[0,1]$; (2) normalizing by $1/K$ to ensure comparability across texts of different lengths; and (3) adaptively selecting $K = \min(\texttt{max\_k}, \max(\texttt{min\_k}, \lfloor \alpha L \rfloor),\ L)$ to balance signal strength, computational efficiency, and stability across short and long sequences.

% -------------------- ASSUMPTION 1 --------------------
\begin{assumption}[Hard-token signal concentration]\normalfont
\label{asm:informative-token}
Let $\delta_i := \Prb[\Delta_i > 0 \mid x \in D_{\mem}] - \Prb[\Delta_i > 0 \mid x \in D_{\non}]$ be the member–non-member gap in the \emph{sign} indicator that \prjctname{} actually aggregates at token~$i$. We assume $\delta_i \ge 0$ and that $\delta_i$ is non-increasing in $p_{\F,i}$, so that harder tokens under~$\mathcal{F}$ carry at least as much membership signal as easier ones, and carry it in the direction the score exploits. The sign constraint matters: a token with $\delta_i < 0$ is discriminative in the opposite direction, and including it would lower the \prjctname{} score for members relative to non-members.
\end{assumption}

\vspace{0.5em}
\noindent\begin{observation}[Informative-token focus]\normalfont
\label{obs:informative-token}
Under Assumption~\ref{asm:informative-token}, the set $I_{\min}$ of the $K$ smallest $p_{\F,i}$ maximizes $\sum_{i \in I} \delta_i$ over all $I$ with $|I| = K$, so restricting the score to low-probability tokens (lines 10-13 of Algorithm~1) retains the most discriminative positions. Because $\delta_i$ is non-negative and non-increasing in $p_{\F,i}$, the $K$ smallest $p_{\F,i}$ are exactly the $K$ largest $\delta_i$, and the claim follows by selecting the largest terms. For a fixed candidate set $I$, this is the quantity the attack depends on: $\E[S \mid x \in D_{\mem}] - \E[S \mid x \in D_{\non}] = \frac{1}{K}\sum_{i \in I}\delta_i$, so maximizing $\sum_{i \in I}\delta_i$ maximizes the mean separation of the deployed score.
\end{observation}

\vspace{0.5em}
\noindent We state Assumption~\ref{asm:informative-token} in terms of the sign gap $\delta_i$ rather than a gap in mean probability differences, because \prjctname{} aggregates the signs $\mathbf{1}\{\Delta_i > 0\}$ and not the raw values $\Delta_i$; a criterion phrased over mean differences would not describe the deployed statistic. The assumption is well-motivated: high-probability tokens are already well-predicted by both models, leaving minimal room for $\mathcal{F}$ to reveal memorization, while low-probability positions are challenging predictions where fine-tuning effects are most pronounced. It is also supported empirically. Figure~\ref{fig:token_prob_improve_cln} shows that member and non-member sign improvements diverge precisely at low-confidence tokens. It remains an assumption about the data rather than a property we verify, so Observation~\ref{obs:informative-token} motivates the design instead of proving it optimal in general.

% -------------------- OBSERVATION 2 --------------------
\begin{observation}[Threshold rule under monotone likelihood ratio]\normalfont
\label{obs:np-optimal}
Let $Z_i := \mathbf{1}\{\Delta_i > 0\}$ for $i \in I_{\min}$ and define $S = \frac{1}{K} \sum_{i \in I_{\min}} Z_i$. Since $S$ averages $K$ binary indicators it is discrete, taking values in $\{0, 1/K, \dots, 1\}$. Suppose $S$ has probability mass functions $p_{\mem}$ and $p_{\non}$ for member and non-member texts whose likelihood ratio $p_{\mem}(s)/p_{\non}(s)$ is monotone increasing in~$s$. Then, among tests based on~$S$ with a fixed false-positive rate, the threshold rule $g(x; \mathcal{F}) = \text{member} \iff S \ge \tau$ is most powerful, by the Neyman–Pearson lemma~\cite{casella2024statistical}.
\end{observation}

\vspace{0.5em}
\noindent Observation~\ref{obs:np-optimal} says that, \emph{where the monotone-likelihood-ratio assumption holds}, thresholding the \prjctname{} score (line 15 of Algorithm~1) is the most powerful rule among those based on~$S$, so practitioners need not search for more complex decision boundaries on the score. Two caveats bound this statement. First, it is optimal only among decisions that use the scalar score $S$, not among all possible attacks. Second, the monotone-likelihood-ratio property is exact when the selected indicators are i.i.d.\ Bernoulli; autoregressive tokens and adaptively selected positions are not independent, so this is an idealisation rather than a property of the setting.

% -------------------- THEOREM --------------------
\begin{theorem}[Finite-sample detection guarantees]\normalfont
\label{thm:hoeffding}
Assume $\{Z_i\}_{i \in I_{\min}}$ are independent and bounded in~$[0,1]$, with $\E[Z_i \mid x \in D_{\mem}] = p_{\mem}$ and $\E[Z_i \mid x \in D_{\non}] = p_{\non}$, where $p_{\mem} > p_{\non}$. For any threshold~$\tau \in (p_{\non}, p_{\mem})$ and any~$K \ge 1$,
{\small
\begin{align*}
\Prb\!\big[\, S \le \tau \,\big|\, x \in D_{\mem} \,\big] &\le \exp\!\big(\!-2K (p_{\mem} - \tau)^2\big), \\[0.3em]
\Prb\!\big[\, S \ge \tau \,\big|\, x \in D_{\non} \,\big] &\le \exp\!\big(\!-2K (\tau - p_{\non})^2\big).
\end{align*}
}
\end{theorem}

\vspace{0.3em}
\noindent\textit{Proof.} The score $S = \frac{1}{K}\sum_{i \in I_{\min}} Z_i$ is the average of $K$ independent variables in $[0,1]$ with $\E[S \mid x \in D_{\mem}] = p_{\mem}$ and $\E[S \mid x \in D_{\non}] = p_{\non}$. Applying Hoeffding's inequality~\cite{hoeffding1963probability} with deviations $t = p_{\mem} - \tau$ and $t = \tau - p_{\non}$ yields the stated bounds.

\vspace{0.5em}
\begin{corollary}[Sample complexity for target power]\normalfont
\label{cor:power}
Let $\gamma := p_{\mem} - p_{\non} > 0$ and set the threshold to the midpoint $\tau = \tfrac{1}{2}(p_{\mem}+p_{\non})$. If $K \ge \frac{2}{\gamma^2} \log\frac{1}{\beta}$, then both the false-positive and false-negative rates are at most $\beta$.
\end{corollary}

\vspace{0.3em}
\noindent\textit{Proof.} Setting each bound in Theorem~\ref{thm:hoeffding} equal to $\beta$ and solving for $K$ gives $K \ge \frac{\log(1/\beta)}{2(p_{\mem} - \tau)^2}$ and $K \ge \frac{\log(1/\beta)}{2(\tau - p_{\non})^2}$. At the midpoint $\tau = \tfrac{1}{2}(p_{\mem}+p_{\non})$ both deviations equal $\gamma/2$, so both bounds reduce to $K \ge \frac{2}{\gamma^2}\log\frac{1}{\beta}$.

\vspace{0.5em}
\noindent Theorem~\ref{thm:hoeffding} provides exponential concentration guarantees: both error types decay exponentially in $K$. Corollary~\ref{cor:power} quantifies how many tokens are needed for reliable detection, and the requirement depends strongly on the gap $\gamma = p_{\mem} - p_{\non}$: at $\beta = 0.05$ a gap of $\gamma = 0.2$ requires $K \ge 150$, while $\gamma = 0.4$ requires only $K \ge 38$. The bound is therefore informative about the scaling rather than a guarantee that the $K$ values we use are sufficient. This motivates our adaptive $K$ selection (lines 10-11 of Algorithm~1): scaling $K$ with sequence length through $\alpha L$, subject to the practical bounds (\texttt{min\_k}, \texttt{max\_k}), lets the number of aggregated tokens grow with the available signal. For longer sequences, larger $K$ provides stronger concentration; for shorter sequences $K$ is capped by the sequence length itself, so \texttt{min\_k} raises the floor on aggregation but cannot on its own ensure the bound is met.

Together, these results motivate the design of \prjctname{} under explicit assumptions: (1) if harder tokens carry at least as much membership signal, selecting low-probability tokens retains the most discriminative positions (Observation~\ref{obs:informative-token}); (2) under a monotone-likelihood-ratio assumption, thresholding the score is most powerful among score-based tests (Observation~\ref{obs:np-optimal}); and (3) given independent selected indicators, the finite-sample error bounds decay exponentially in the number of selected tokens $K$ (Theorem~\ref{thm:hoeffding}). Observation~\ref{obs:informative-token} rests on the hard-token monotonicity of Assumption~\ref{asm:informative-token}, Observation~\ref{obs:np-optimal} on the monotone-likelihood-ratio property, and Theorem~\ref{thm:hoeffding} on the independence of the selected indicators. These assumptions are idealisations of autoregressive decoding and adaptive token selection, and none holds strictly in our setting; the results therefore provide conditional intuition rather than a universal optimality guarantee. In practice the method remains effective even when the reference model $\mathcal{R}$ is not exactly the pre-fine-tuning checkpoint.

\section{Experimental Setup}
\label{sec:appendix_setup}
Table~\ref{tab:finetune_config} summarizes the hyperparameters used to fine-tune each target model. All models are fully fine-tuned with seed $42$; sequences are truncated and right-padded to $512$ tokens, one example per sequence, with no document packing.
For HT-MIA, adaptive token selection uses
$\text{min\_k}=20$, $\text{max\_k}=30$, and $\alpha = 0.2$, with $K = \min(\text{max\_k},\,
\max(\text{min\_k},\ \lfloor \alpha L \rfloor),\ L)$ where $L$ is the input sequence
length. These values are fixed for every dataset and target model, with no per-dataset tuning.
% $\beta=0.5$, and a smoothing factor of $\varepsilon=10^{-7}$ on 
% token probabilities to avoid numerical instability.
All experiments were conducted on a server with 7 NVIDIA H200 GPUs. Fine-tuning of LLaMA-3.1-70B is sharded across six of these GPUs, whereas its evaluation runs on a single GPU with 8-bit weights; Appendix~\ref{sec:appendix_scaling} gives the full configuration.

\begin{table}[h]
\centering
\small
\setlength{\tabcolsep}{1pt}
\resizebox{\columnwidth}{!}{%
\begin{tabular}{lcccccc}
\toprule
\textbf{Parameter} & \textbf{GPT-2} & \textbf{Qwen3-0.6B} & \textbf{LLaMA-3.2-1B} & \textbf{LLaMA-3.1-8B} & \textbf{LLaMA-3.1-70B} \\
\midrule
Optimizer & AdamW & AdamW & AdamW & 8-bit AdamW & 8-bit AdamW \\
LR           & $1\times10^{-4}$ & $1\times10^{-4}$ & $1\times10^{-4}$ & $2\times10^{-5}$ & $2\times10^{-5}$ \\
Batch Size & 8 & 8 & 8 & 4 & 2 \\
Grad. Accum. Steps    & 1 & 2 & 2 & 2 & 4 \\
Epochs                  & 1 & 1 & 1 & 1 & 1 \\
Max Seq. Len.          & 512 & 512 & 512 & 512 & 512 \\
Weight Decay            & 0.01 & 0.01 & 0.01 & 0.01 & 0.01 \\
Max Grad Norm           & 1.0 & 1.0 & 1.0 & 1.0 & 1.0 \\
LR Scheduler            & cosine & cosine & cosine & cosine & cosine \\
Warmup                  & 10\% steps & 10\% steps & 10\% steps & 10\% steps & 10\% steps \\
Precision               & fp32 & bf16 & bf16 & bf16 & bf16 \\
\bottomrule
\end{tabular}
}
\caption{Fine-tuning hyperparameters for each model.}
\label{tab:finetune_config}
\end{table}

\section{Baselines}
\label{sec:appendix_B}
To compare \prjctname{}, we evaluate against seven established MIA attack methods.  
  
\textbf{Loss.}
The \textit{Loss}-based attack~\cite{yeom2018privacy} infers membership by using the per-sample cross-entropy loss, under the assumption that training samples generally achieve lower loss values than non-members:
\begin{equation}
\setlength\abovedisplayskip{1pt}
\setlength\belowdisplayskip{1pt}
% \small
L(x, y) = - \log p_\theta(y \mid x),
\end{equation}
where $x$ denotes the input text, $y$ is the target output, and $p_\theta(y \mid x)$ represents the model's predicted probability for $y$ given $x$ under parameters $\theta$. Lower loss values suggest the model is more confident in its predictions, which is typical for training samples, making $L(x, y)$ an effective membership signal. Samples with $L(x, y)$ below a chosen threshold are classified as members.

\textbf{Lowercase.}
The \textit{Lowercase} attack~\cite{carlini2021extracting} evaluates memorization by measuring the change in model perplexity when the input text is normalized to lowercase form:
\begin{equation}
\setlength\abovedisplayskip{1pt}
\setlength\belowdisplayskip{1pt}
% \small
\Delta \text{PPL} = \text{PPL}(x) - \text{PPL}(\text{lowercase}(x)),
\end{equation}
where $\text{PPL}(x) = \exp(L(x))$ is the perplexity of the original text computed as the exponential of the average per-token negative log-likelihood, and $\text{lowercase}(x)$ denotes the text with all characters converted to lowercase. Memorized sequences are typically more sensitive to this perturbation, producing larger changes in perplexity ($\Delta \text{PPL} > 0$) compared to non-members, which exhibit more robust predictions.

% \textbf{Min-K and Min-K++.}
\textbf{Min-K++.}
% The \textit{Min-K}~\cite{shi2023detecting} method identifies membership by focusing on the least likely $k\%$ tokens in a sequence:
% \begin{equation}
% \setlength\abovedisplayskip{1pt}
% \setlength\belowdisplayskip{1pt}
% % \small
% \text{Min-K\%}(x) = \frac{1}{|k\%|} \sum_{t \in \text{min-}k\%} -\log p_\theta(x_t \mid x_{<t}),
% \end{equation}
% where $x_t$ is the token at position $t$, $x_{<t}$ represents all preceding tokens, $p_\theta(x_t \mid x_{<t})$ is the model's predicted probability for token $x_t$ given its context, and $\text{min-}k\%$ denotes the set of token positions corresponding to the lowest $k\%$ of probabilities in the sequence. The term $|k\%|$ denotes the number of tokens selected. The intuition is that member texts exhibit lower negative log-likelihoods even at their most difficult tokens compared to non-members.
% The \textit{Min-K++}~\cite{zhang2024min} method extends this idea by normalizing token probabilities to achieve more stable discrimination:
% \begin{equation}
% \setlength\abovedisplayskip{1pt}
% \setlength\belowdisplayskip{1pt}
% \small
% \text{Min-K\%++}(x) = \frac{1}{|k\%|} \sum_{t \in \text{min-}k\%} \frac{\log p_\theta(x_t \mid x_{<t}) - \mu}{\sigma},
% \end{equation}
% where $\mu$ and $\sigma$ are the mean and standard deviation of log-probabilities computed over all tokens in the sequence $x$. This z-score normalization accounts for per-sequence variability in token difficulty, making the score more robust across texts with different intrinsic complexity levels. Higher scores indicate membership.
The \textit{Min-K++}~\cite{zhang2024min} method identifies membership by focusing on the least likely $k\%$ tokens in a sequence, normalizing each token against the model's next-token distribution at that same position:
\begin{equation}
\setlength\abovedisplayskip{1pt}
\setlength\belowdisplayskip{1pt}
\begin{aligned}
\text{Min-K\%++}(x) &= \frac{1}{|k\%|} \sum_{t \in \text{min-}k\%} z_t, \\[0.3em]
z_t &= \frac{\log p_\theta(x_t \mid x_{<t}) - \mu_t}{\sigma_t},
\end{aligned}
\end{equation}
where $x_t$ is the token at position $t$, $x_{<t}$ represents all preceding tokens, $p_\theta(x_t \mid x_{<t})$ is the model's predicted probability for token $x_t$ given its context, and $\text{min-}k\%$ denotes the set of token positions corresponding to the lowest $k\%$ of probabilities in the sequence. The term $|k\%|$ denotes the number of tokens selected. Crucially, the normalization is \emph{position-specific}: $\mu_t$ and $\sigma_t$ are the mean and standard deviation of the log-probabilities $\{\log p_\theta(v \mid x_{<t})\}_{v \in \mathcal{V}}$ that the model assigns across the entire vocabulary $\mathcal{V}$ at position $t$, rather than a single pair of statistics shared across the sequence. Each token is therefore judged by how unlikely it is relative to the alternatives available at its own position, which calibrates the score against the local uncertainty of the model. Higher scores indicate membership.

\textbf{PAC.}
The \textit{PAC} attack~\cite{ye2024data} contrasts a polarized token–log-probability distance on the original text with the average over augmented neighbors $\mathcal{A}(x)$:
\begin{equation}
\setlength\abovedisplayskip{1pt}
\setlength\belowdisplayskip{1pt}
% \small
\mathrm{PAC}(x) \;=\; \mathrm{PD}(x) \;-\; \tfrac{1}{|\mathcal{A}(x)|}\!\sum_{\tilde x\in\mathcal{A}(x)} \mathrm{PD}(\tilde x),
\end{equation}
where $\mathrm{PD}(x)$ denotes the \textit{polarized distance}, defined as the difference between the mean log-probabilities of the top-$k$ and bottom-$k$ tokens in sequence $x$. The set $\mathcal{A}(x)$ contains augmented versions of $x$ generated through paraphrasing or perturbation operations, and $|\mathcal{A}(x)|$ is the number of such augmentations. The second term represents the average polarized distance over these neighbors. The intuition is that member texts exhibit a larger gap between easy and hard tokens compared to their augmented variants, as memorization creates sharp probability contrasts that are disrupted by text modification.

\textbf{Ratio.}
The \textit{Ratio} attack~\cite{carlini2021extracting} compares the per-sample negative log-likelihood (NLL) under the target model to that under a fixed reference model:
\begin{equation}
\setlength\abovedisplayskip{1pt}
\setlength\belowdisplayskip{1pt}
% \small
\mathrm{Ratio}(x) \;=\; -\,\frac{\mathrm{NLL}_{\text{tgt}}(x)}{\mathrm{NLL}_{\text{ref}}(x)} ,
\end{equation}
where $\mathrm{NLL}_{\text{tgt}}(x) = -\sum_t \log p_{\theta_{\text{tgt}}}(x_t \mid x_{<t})$ is the negative log-likelihood under the target (fine-tuned) model with parameters $\theta_{\text{tgt}}$, and $\mathrm{NLL}_{\text{ref}}(x)$ is the corresponding quantity under a reference model with parameters $\theta_{\text{ref}}$. The negative sign ensures that larger ratios, which indicate lower loss on the target model relative to the reference, correspond to higher membership likelihood. This method captures relative improvement: members should have lower NLL under the target model compared to the baseline reference.

\textbf{Zlib.}
The \textit{Zlib}-based attack~\cite{carlini2021extracting} contrasts model fit with text compressibility:
\begin{equation}
\setlength\abovedisplayskip{1pt}
\setlength\belowdisplayskip{1pt}
% \small
\mathrm{Zlib}(x) \;=\; -\,\frac{L(x)}{\mathrm{zlibEntropy}(x)} ,
\end{equation}
where $L(x) = -\frac{1}{|x|}\sum_t \log p_\theta(x_t \mid x_{<t})$ is the per-token average NLL under the model, and $\mathrm{zlibEntropy}(x) = \frac{\mathrm{len}(\mathrm{zlib.compress}(x))}{\mathrm{len}(x)}$ is the compressed size per character obtained by applying the zlib compression algorithm to the text. The intuition is that memorized texts have low model loss $L(x)$ relative  to their intrinsic complexity measured by compression. A small loss-to-entropy ratio, and hence a larger score, suggests the model fits the text better than its complexity warrants, indicating memorization. Samples with larger $\mathrm{Zlib}(x)$ scores are classified as members.

\textbf{SPV-MIA.}
The \textit{SPV-MIA} (Self-calibrated Probabilistic Variation) attack~\cite{fu2023practical} measures probabilistic variation by comparing model probabilities on paraphrased versus original text:
\begin{equation}
\setlength\abovedisplayskip{1pt}
\setlength\belowdisplayskip{1pt}
\mathrm{SPV}(x) \;=\; \left[\bar{p}_{\theta}(\tilde{x}) - p_{\theta}(x)\right] - \left[\bar{p}_{\hat{\theta}}(\tilde{x}) - p_{\hat{\theta}}(x)\right] ,
\end{equation}
where $\bar{p}_{\theta}(\tilde{x}) = \frac{1}{N}\sum_{i=1}^N p_\theta(\tilde{x}_i)$ is the mean probability assigned by the target model with parameters $\theta$ to $N$ paraphrased versions $\tilde{x}_i$ generated via T5-based mask-filling, $p_{\theta}(x)$ is the probability on the original text. $\bar{p}_{\hat{\theta}}(\tilde{x})$, $p_{\hat{\theta}}(x)$ are the corresponding quantities on a self-prompt reference model with parameters $\hat{\theta}$. The reference model is fine-tuned on text generated by prompting the target model itself, enabling calibration without access to the original training data. The intuition is that memorized texts exhibit smaller probability variation across paraphrases on the target model compared to the reference model. Because member texts resist such perturbations due to memorization, they yield higher calibrated scores. Samples with more positive $\mathrm{SPV}(x)$ values are classified as members.

\section{Evaluation Metrics}
\label{sec:appendix_C}
To evaluate the effectiveness of \prjctname{}, we report the Area Under the ROC Curve (AUC) and the True Positive Rate (TPR) at fixed False Positive Rate (FPR)~\cite{carlini2022membership}.
The outcomes of the attack classifier are summarized by the standard quantities of true positives (TP), false positives (FP), true negatives (TN), and false negatives (FN).
The evaluation sets follow the 50:50 member/non-member split described in the experimental setup. Asclepius has $79{,}057$ members and $79{,}057$ non-members, IMDB $27{,}500$ each, Clinicalnotes $15{,}000$ each, and Wikipedia $5{,}000$ each. These sizes set the resolution at strict operating points. At FPR$=0.001$ they allow roughly $79$, $28$, $15$ and $5$ false positives respectively. The reported TPR values at that setting are therefore finite-sample estimates, and small differences between methods should not be over-interpreted.

% \paragraph{True Positive Rate (TPR).}  
\textbf{True Positive Rate (TPR).}
The TPR measures the fraction of member samples correctly classified as members:
\begin{equation}
\setlength\abovedisplayskip{1pt}
\setlength\belowdisplayskip{1pt}
% \small
\text{TPR} = \frac{TP}{TP + FN}.
\end{equation}

% \paragraph{False Positive Rate (FPR).} 
\textbf{False Positive Rate (FPR).}
The FPR measures the fraction of non-member samples incorrectly classified as members:
\begin{equation}
\setlength\abovedisplayskip{1pt}
\setlength\belowdisplayskip{1pt}
% \small
\text{FPR} = \frac{FP}{FP + TN}.
\end{equation}
In our evaluation, we report TPR@FPR=0.1 and TPR@FPR=0.01, which quantify the attack’s detection power under strict false-positive constraints.  

\textbf{ROC Curve and Area Under the Curve (AUC).}
The ROC curve plots TPR against FPR across varying decision thresholds of the attack classifier. The area under the ROC curve (AUC) provides a threshold-independent measure of discrimination power:
\begin{equation}
\setlength\abovedisplayskip{1pt}
\setlength\belowdisplayskip{1pt}
% \small
\text{AUC} = \int_0^1 \text{TPR}(\text{FPR}) \, d(\text{FPR}).
\end{equation}
AUC values range from 0.5 (random guessing) to 1.0 (perfect classification), with higher values indicating better MIA performance.

\textbf{Binary Classification Accuracy.}
We additionally report binary classification accuracy, referred to simply as accuracy, at a fixed decision threshold $\tau$, where a sample is predicted as a member if the attack score exceeds $\tau$ and as a non-member otherwise:
\begin{equation}
\setlength\abovedisplayskip{1pt}
\setlength\belowdisplayskip{1pt}
\text{ACC} = \frac{TP + TN}{TP + TN + FP + FN}.
\end{equation}

\textbf{Perplexity (PPL).} For the privacy analysis we also report per-sample perplexity,
\begin{equation}
\setlength\abovedisplayskip{1pt}
\setlength\belowdisplayskip{1pt}
\text{PPL}(x) = \exp\!\Big(\tfrac{1}{n}\sum_{i=1}^{n} -\log p_\theta(x_i \mid x_{<i})\Big),
\end{equation}
the exponential of the average per-token negative log-likelihood; lower PPL indicates a better fit to the text.

\section{\prjctname{} vs.\ Baselines on General Datasets}
\label{sec:appendix_genData}
Tables~\ref{tab:allmetrics_imdb} and~\ref{tab:allmetrics_wiki} highlight the performance 
of \prjctname{} against all baselines across IMDB and Wikipedia datasets, respectively.

\begin{table*}[!ht]
\centering
\setlength{\tabcolsep}{6pt}
\begin{subtable}[t]{0.32\textwidth}
\centering
\resizebox{\textwidth}{!}{%
\begin{tabular}{l c c}
    \toprule
    \textbf{Method} & \textbf{LLaMA-3.2-1B} & \textbf{Qwen-3-0.6B} \\
    \midrule
    Loss         & 0.8977 & 0.8939 \\
    Lowercase    & 0.8237 & 0.8132 \\
    Min-K++      & 0.8286 & 0.8160 \\
    PAC          & 0.8868 & 0.8812 \\
    \textbf{Ratio}        & \textbf{0.9593} & \textbf{0.9741} \\
    Zlib         & 0.8995 & 0.8981 \\
    SPV-MIA      & 0.9101 & 0.9329 \\
    \prjctname{} & 0.9225 & 0.9468 \\
    \bottomrule
\end{tabular}}
\caption{AUC}
\label{tab:auc_llama_qwen_imdb}
\end{subtable}
\hfill
\begin{subtable}[t]{0.32\textwidth}
\centering
\resizebox{\textwidth}{!}{%
\begin{tabular}{l c c}
    \toprule
    \textbf{Method} & \textbf{LLaMA-3.2-1B} & \textbf{Qwen-3-0.6B} \\
    \midrule
    Loss         & 0.7164 & 0.6947 \\
    Lowercase    & 0.5442 & 0.4786 \\
    Min-K++      & 0.5733 & 0.5064 \\
    PAC          & 0.6815 & 0.6768 \\
    \textbf{Ratio}        & \textbf{0.8604} & \textbf{0.9209} \\
    Zlib         & 0.7226 & 0.7085 \\
    SPV-MIA      & 0.7210 & 0.7930 \\
    \prjctname{} & 0.7895 & 0.8524 \\
    \bottomrule
\end{tabular}}
\caption{TPR@FPR=0.1}
\label{tab:tpr01_llama_qwen_imdb}
\end{subtable}
\hfill
\begin{subtable}[t]{0.32\textwidth}
\centering
\resizebox{\textwidth}{!}{%
\begin{tabular}{l c c}
    \toprule
    \textbf{Method} & \textbf{LLaMA-3.2-1B} & \textbf{Qwen-3-0.6B} \\
    \midrule
    Loss         & 0.3744 & 0.2674 \\
    Lowercase    & 0.1941 & 0.1449 \\
    Min-K++      & 0.1893 & 0.1274 \\
    PAC          & 0.3221 & 0.2668 \\
    \textbf{Ratio}        & \textbf{0.6838} & \textbf{0.7723} \\
    Zlib         & 0.4004 & 0.2932 \\
    SPV-MIA      & 0.2240 & 0.3500 \\
    \prjctname{} & 0.5984 & 0.6098 \\
    \bottomrule
\end{tabular}}
\caption{TPR@FPR=0.01}
\label{tab:tpr001_llama_qwen_imdb}
\end{subtable}
\caption{Comparison of AUC, TPR@FPR=0.1, and TPR@FPR=0.01 across LLaMA-3.2-1B and Qwen-3-0.6B fine-tuned on the IMDB dataset.}
\label{tab:allmetrics_imdb}
\end{table*}

\begin{table*}[!ht]
\centering
\setlength{\tabcolsep}{6pt}
\begin{subtable}[t]{0.32\textwidth}
\centering
\resizebox{\textwidth}{!}{%
\begin{tabular}{l c c}
    \toprule
    \textbf{Method} & \textbf{LLaMA-3.2-1B} & \textbf{Qwen-3-0.6B} \\
    \midrule
    Loss         & 0.7574 & 0.7553 \\
    Lowercase    & 0.6433 & 0.6524 \\
    Min-K++      & 0.7639 & 0.7734 \\
    PAC          & 0.7354 & 0.7616 \\
    Ratio        & 0.8153 & 0.8372 \\
    Zlib         & 0.7314 & 0.7302 \\
    SPV-MIA      & 0.7889 & 0.8065 \\
    \textbf{\prjctname{}} & \textbf{0.8621} & \textbf{0.9100} \\
    \bottomrule
\end{tabular}}
\caption{AUC}
\label{tab:auc_llama_qwen_wiki}
\end{subtable}
\hfill
\begin{subtable}[t]{0.32\textwidth}
\centering
\resizebox{\textwidth}{!}{%
\begin{tabular}{l c c}
    \toprule
    \textbf{Method} & \textbf{LLaMA-3.2-1B} & \textbf{Qwen-3-0.6B} \\
    \midrule
    Loss         & 0.4104 & 0.3152 \\
    Lowercase    & 0.1834 & 0.1932 \\
    Min-K++      & 0.4292 & 0.3950 \\
    PAC          & 0.3300 & 0.3396 \\
    Ratio        & 0.5756 & 0.3936 \\
    Zlib         & 0.3280 & 0.2564 \\
    SPV-MIA      & 0.4080 & 0.5700 \\
    \textbf{\prjctname{}} & \textbf{0.7158} & \textbf{0.8020} \\
    \bottomrule
\end{tabular}}
\caption{TPR@FPR=0.1}
\label{tab:tpr01_llama_qwen_wiki}
\end{subtable}
\hfill
\begin{subtable}[t]{0.32\textwidth}
\centering
\resizebox{\textwidth}{!}{%
\begin{tabular}{l c c}
    \toprule
    \textbf{Method} & \textbf{LLaMA-3.2-1B} & \textbf{Qwen-3-0.6B} \\
    \midrule
    Loss         & 0.0450 & 0.0392 \\
    Lowercase    & 0.0170 & 0.0158 \\
    Min-K++      & 0.0762 & 0.0538 \\
    PAC          & 0.0216 & 0.0236 \\
    Ratio        & 0.0456 & 0.0184 \\
    Zlib         & 0.0254 & 0.0246 \\
    SPV-MIA      & 0.0350 & 0.0530 \\
    \textbf{\prjctname{}} & \textbf{0.5550} & \textbf{0.5002} \\
    \bottomrule
\end{tabular}}
\caption{TPR@FPR=0.01}
\label{tab:tpr001_llama_qwen_wiki}
\end{subtable}
\caption{Comparison of AUC, TPR@FPR=0.1, and TPR@FPR=0.01 across LLaMA-3.2-1B and Qwen-3-0.6B fine-tuned on the Wikipedia dataset.}
\label{tab:allmetrics_wiki}
\end{table*}

On the IMDB dataset, the results present a notable departure from previous patterns. 
Ratio emerges as the top-performing method on both LLaMA-3.2-1B and Qwen-3-0.6B 
(Table~\ref{tab:allmetrics_imdb}), achieving AUCs of 95.9\% and 97.4\%, while
\prjctname{} attains 92.3\% and 94.7\%, the only case where our method does not
achieve the highest AUC. Multiple baselines also perform substantially better on IMDB, with SPV-MIA, Zlib, and Loss all exceeding 89\%. The convergence of methods at high performance levels ($>$88\%) suggests that IMDB's short, sentiment-driven reviews produce strong membership signals that diminish observable differences among competing approaches. Despite not achieving the highest overall AUC, \prjctname{} remains the second-best method across all metrics, achieving TPR@FPR=0.01 of 59.8\% and 61.0\% on LLaMA-3.2-1B and Qwen-3-0.6B respectively, approximately 1.5$\times$ and 2.1$\times$ the third-best method (Zlib at 40.0\% and 29.3\%). On the Wikipedia dataset (Table~\ref{tab:allmetrics_wiki}), \prjctname{} achieves AUCs of 86.2\% and 91.0\% and TPR@FPR=0.01 of 55.5\% and 50.0\% on LLaMA-3.2-1B and Qwen-3-0.6B, outperforming all baselines by a large margin.
Table~\ref{tab:allmetrics_wiki} further confirms this trend, with \prjctname{} 
demonstrating even stronger dominance on the Wikipedia dataset under strict 
false positive constraints.

\section{Comparison with State-of-the-Art MIA Methods}
\label{sec:appendix_sota}
Table~\ref{tab:sota_comparison} presents a comprehensive comparison of \prjctname{} against two state-of-the-art MIA methods, EZ-MIA and WBC across four datasets and three target models.

\begin{table*}[t]
\centering
\small
\begin{tabular}{llcccc}
\toprule
\textbf{Model} & \textbf{Method} & \textbf{IMDB} & \textbf{Wikipedia} & \textbf{Clinicalnotes} & \textbf{Asclepius} \\
\midrule

\multicolumn{6}c{\textbf{(a) AUC}} \\
\midrule
\multirow{3}{*}{\textit{Qwen3-0.6B}} 
  & EZ-MIA & \textbf{0.9762} & \textbf{0.9257} & \textbf{0.8980} & 0.6703 \\
  & WBC    & 0.9650          & 0.9129          & 0.8809          & \textbf{0.7584} \\
  & HT-MIA & 0.9468          & 0.9100          & 0.8843          & 0.7312 \\
\addlinespace
\multirow{3}{*}{\textit{LLaMA-3.2-1B}} 
  & EZ-MIA & \textbf{0.9599} & \textbf{0.8735} & 0.8423          & \textbf{0.7787} \\
  & WBC    & 0.9434          & 0.8663          & 0.8594          & 0.7619 \\
  & HT-MIA & 0.9225          & 0.8621          & \textbf{0.8659} & 0.7348 \\
\addlinespace
\multirow{3}{*}{\textit{GPT-2}} 
  & EZ-MIA & \textbf{0.9423} & 0.7732          & \textbf{0.7128} & 0.5738 \\
  & WBC    & 0.9296          & \textbf{0.7957} & 0.7122          & \textbf{0.5769} \\
  & HT-MIA & 0.9009          & 0.7448          & 0.6978          & 0.5679 \\

\midrule
\multicolumn{6}c{\textbf{(b) TPR@FPR=0.1}} \\
\midrule
\multirow{3}{*}{\textit{Qwen3-0.6B}} 
  & EZ-MIA & \textbf{0.9308} & \textbf{0.8386} & \textbf{0.7683} & 0.3297 \\
  & WBC    & 0.9121          & 0.8101          & 0.7090          & \textbf{0.3873} \\
  & HT-MIA & 0.8524          & 0.8020          & 0.6861          & 0.3129 \\
\addlinespace
\multirow{3}{*}{\textit{LLaMA-3.2-1B}} 
  & EZ-MIA & \textbf{0.8720} & \textbf{0.7380} & 0.6990          & \textbf{0.5051} \\
  & WBC    & 0.8544          & 0.7250          & \textbf{0.7153} & 0.4374 \\
  & HT-MIA & 0.7895          & 0.7158          & 0.7057          & 0.3707 \\
\addlinespace
\multirow{3}{*}{\textit{GPT-2}} 
  & EZ-MIA & 0.8028          & 0.3920          & \textbf{0.3380} & 0.1473 \\
  & WBC    & \textbf{0.8273} & \textbf{0.4373} & 0.3293          & \textbf{0.1580} \\
  & HT-MIA & 0.7139          & 0.3622          & 0.3223          & 0.1109 \\

\midrule
\multicolumn{6}c{\textbf{(c) TPR@FPR=0.01}} \\
\midrule
\multirow{3}{*}{\textit{Qwen3-0.6B}} 
  & EZ-MIA & 0.7945          & \textbf{0.6146} & \textbf{0.4297} & 0.0706 \\
  & WBC    & \textbf{0.8032} & 0.4036          & 0.3766          & \textbf{0.0780} \\
  & HT-MIA & 0.6098          & 0.5002          & 0.3839          & 0.0477 \\
\addlinespace
\multirow{3}{*}{\textit{LLaMA-3.2-1B}} 
  & EZ-MIA & \textbf{0.7318} & \textbf{0.5832} & \textbf{0.4691} & \textbf{0.1224} \\
  & WBC    & 0.7180          & 0.5485          & 0.4198          & 0.0963 \\
  & HT-MIA & 0.5984          & 0.5550          & 0.4253          & 0.0817 \\
\addlinespace
\multirow{3}{*}{\textit{GPT-2}} 
  & EZ-MIA & 0.4568          & 0.0458          & 0.0560          & \textbf{0.0240} \\
  & WBC    & \textbf{0.4913} & 0.0527          & \textbf{0.0667} & 0.0200 \\
  & HT-MIA & 0.2905          & \textbf{0.0696} & 0.0524          & 0.0181 \\

\midrule
\multicolumn{6}c{\textbf{(d) TPR@FPR=0.001}} \\
\midrule
\multirow{3}{*}{\textit{Qwen3-0.6B}}
  & EZ-MIA & \textbf{0.6953} & \textbf{0.2893} & \textbf{0.1947} & \textbf{0.0140} \\
  & WBC    & 0.6480          & 0.1220          & 0.1707          & 0.0127 \\
  & HT-MIA & 0.3476          & 0.2020          & 0.1470          & 0.0134 \\
\addlinespace
\multirow{3}{*}{\textit{LLaMA-3.2-1B}}
  & EZ-MIA & 0.6320          & \textbf{0.4673} & \textbf{0.2373} & 0.0080 \\
  & WBC    & \textbf{0.6433} & 0.3473          & 0.1353          & 0.0073 \\
  & HT-MIA & 0.5166          & 0.2206          & 0.1610          & \textbf{0.0143} \\
\addlinespace
\multirow{3}{*}{\textit{GPT-2}}
  & EZ-MIA & 0.2178          & \textbf{0.0040} & --              & -- \\
  & WBC    & \textbf{0.2987} & 0.0027          & 0.0093          & \textbf{0.0020} \\
  & HT-MIA & 0.0804          & 0.0026          & \textbf{0.0143} & 0.0013 \\
\bottomrule
\end{tabular}
\caption{Comparison of HT-MIA against state-of-the-art MIA methods (EZ-MIA and WBC). -- denotes an unavailable result; no cross-method comparison is made for that cell. At FPR$=0.001$ the reported values are finite-sample estimates. Appendix~\ref{sec:appendix_C} gives the evaluation set sizes.}
\label{tab:sota_comparison}
\end{table*}

As shown in Table~\ref{tab:sota_comparison}, \prjctname{} is competitive with EZ-MIA and WBC in AUC on several settings and achieves the best AUC on Clinicalnotes with LLaMA-3.2-1B. On other datasets, and at stricter false-positive rates, it often trails these newer methods. EZ-MIA discards correctly-predicted positions and WBC aggregates over binary windows, whereas \prjctname{} scores the least-confident tokens directly; we regard these as structural differences rather than proven disadvantages. The appeal of \prjctname{} lies in its simplicity and two-forward-pass computation. We regard it as a novel perspective for performing effective membership inference attacks, which is also the main contribution by \prjctname{}.

\section{Scaling Analysis Across LLM Sizes}
\label{sec:appendix_scaling}
We evaluate \prjctname{} against MIA baselines, including the strong
reference-based Ratio attack, on Llama 3.1 8B and 70B
using the Asclepius dataset (Table~\ref{tab:scaling}). \prjctname{}
achieves the best AUC and TPR@FPR=0.1 at both scales, and the best
TPR@FPR=0.01 on Llama 3.1 8B. On Llama 3.1 8B, it attains an AUC of
74.12\% and a TPR@FPR=0.01 of 12.42\%, more than $2\times$ the next
best baseline. On Llama 3.1 70B, AUC improves further to 77.40\%,
showing that the hard-token signal persists at 8B and 70B scale in the
evaluated Asclepius setting. The one
exception is TPR@FPR=0.01 on Llama 3.1 70B, where the Ratio attack
(12.20\%) edges out \prjctname{} (6.64\%); \prjctname{} nonetheless
remains the strongest method compared to the standard baselines on the
remaining five of the six metric--scale combinations.

Both scaling runs use full fine-tuning; no adapter or low-rank parameterisation is involved. The model is sharded layer-wise across GPUs within a single process, with gradient checkpointing enabled, rather than through FSDP, DeepSpeed or DDP; the 70B run occupies six of the seven H200 GPUs. Quantisation enters in two distinct places and should not be conflated. During fine-tuning, only the optimiser states are quantised (8-bit AdamW), while the model weights remain in bf16 throughout. Separately, at attack time the 70B model is loaded with 8-bit weights; this applies to the 70B evaluation only, and both the target and the reference model are quantised identically, so the token-level comparison is not distorted by asymmetric precision between the two.

\begin{table*}[!ht]
\centering
\setlength{\tabcolsep}{6pt}
\begin{subtable}[t]{0.32\textwidth}
\centering
\resizebox{\textwidth}{!}{%
\begin{tabular}{l c c}
    \toprule
    \textbf{Method} & \textbf{Llama-3.1-8B} & \textbf{Llama-3.1-70B} \\
    \midrule
    Loss         & 0.6387 & 0.6580 \\
    Lowercase    & 0.6508 & 0.6229 \\
    Min-K++      & 0.6056 & 0.6324 \\
    PAC          & 0.6296 & 0.6734 \\
    Zlib         & 0.6809 & 0.6810 \\
    SPV-MIA      & 0.6667 & 0.7139 \\
    Ratio        & 0.6836 & 0.7326 \\
    \textbf{\prjctname{}} & \textbf{0.7412} & \textbf{0.7740} \\
    \bottomrule
\end{tabular}}
\caption{AUC}
\label{tab:auc_scaling}
\end{subtable}
\hfill
\begin{subtable}[t]{0.32\textwidth}
\centering
\resizebox{\textwidth}{!}{%
\begin{tabular}{l c c}
    \toprule
    \textbf{Method} & \textbf{Llama-3.1-8B} & \textbf{Llama-3.1-70B} \\
    \midrule
    Loss         & 0.2477 & 0.2740 \\
    Lowercase    & 0.2749 & 0.2039 \\
    Min-K++      & 0.1966 & 0.2109 \\
    PAC          & 0.2011 & 0.2785 \\
    Zlib         & 0.2979 & 0.2986 \\
    SPV-MIA      & 0.2610 & 0.2770 \\
    Ratio        & 0.2726 & 0.3916 \\
    \textbf{\prjctname{}} & \textbf{0.4136} & \textbf{0.4077} \\
    \bottomrule
\end{tabular}}
\caption{TPR@FPR=0.1}
\label{tab:tpr01_scaling}
\end{subtable}
\hfill
\begin{subtable}[t]{0.32\textwidth}
\centering
\resizebox{\textwidth}{!}{%
\begin{tabular}{l c c}
    \toprule
    \textbf{Method} & \textbf{Llama-3.1-8B} & \textbf{Llama-3.1-70B} \\
    \midrule
    Loss         & 0.0405 & 0.0516 \\
    Lowercase    & 0.0608 & 0.0352 \\
    Min-K++      & 0.0342 & 0.0430 \\
    PAC          & 0.0309 & 0.0583 \\
    Zlib         & 0.0542 & 0.0542 \\
    SPV-MIA      & 0.0370 & 0.0400 \\
    Ratio        & 0.0492 & \textbf{0.1220} \\
    \textbf{\prjctname{}} & \textbf{0.1242} & 0.0664 \\
    \bottomrule
\end{tabular}}
\caption{TPR@FPR=0.01}
\label{tab:tpr001_scaling}
\end{subtable}
\caption{MIA performance of \prjctname{} and baselines on Llama 3.1 8B and 70B (Asclepius dataset). EZ-MIA and WBC are compared against \prjctname{} in Table~\ref{tab:sota_comparison} and are omitted here to keep the scaling study focused on standard baselines. The 70B model is evaluated with 8-bit quantization on a single GPU.}
\label{tab:scaling}
\end{table*}

% \section{Ablation Studies} 
% \label{subsec:abltn}
% % \subsection{Effectiveness of \prjctname{}}
% In this section, we investigate two key questions: (1) why it is necessary to measure performance improvements at low-confidence (hard) tokens rather than aggregating over entire inputs, and (2) why traditional membership inference methods fail.

% Figure~\ref{fig:token_prob_improve_imdb, token_prob_improve_imdb_cln} shows a member and non-member sample from Clinicalnotes with probabilities from both target and reference models obtained from Qwen-3-0.6B. Both samples show probability improvements from reference to target model, even for non-members (Figures~\ref{fig:member_sample} and~\ref{fig:nonmember_sample}). However, the magnitude differs: member improvements exceed 0.2, while non-member improvements barely exceed 0.1. More critically, the most difficult tokens (shown in green) reveal a pronounced distinction: about 91\% improve in member samples versus only 70\% in non-members. This fine-grained signal is diluted when aggregating over entire sequences, explaining why traditional methods like Loss-based or Zlib MIA fail. By focusing on the low-confidence (hard-to-predict) tokens, \prjctname{} captures a much stronger membership signal. 

% We provide further analysis in editing.
% Appendinx~\ref{sec:appendix_D}, \ref{sec:appendix_E}, and \ref{RefModelxC}  

\begin{figure*}[!ht]
  \centering
  \footnotesize
  \begin{subfigure}{1\textwidth}
    \includegraphics[width=\linewidth]{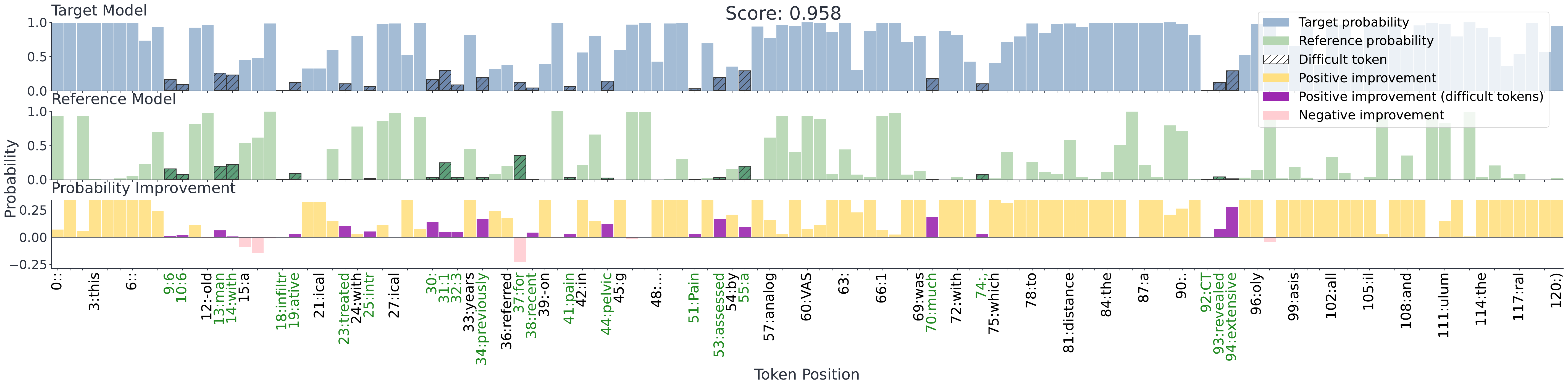}
    \caption{Member sample}
    \label{fig:member_sample}
  \end{subfigure}
  \vspace{4mm} % space between the two subfigures
  \begin{subfigure}{1\textwidth}
    \includegraphics[width=\linewidth]{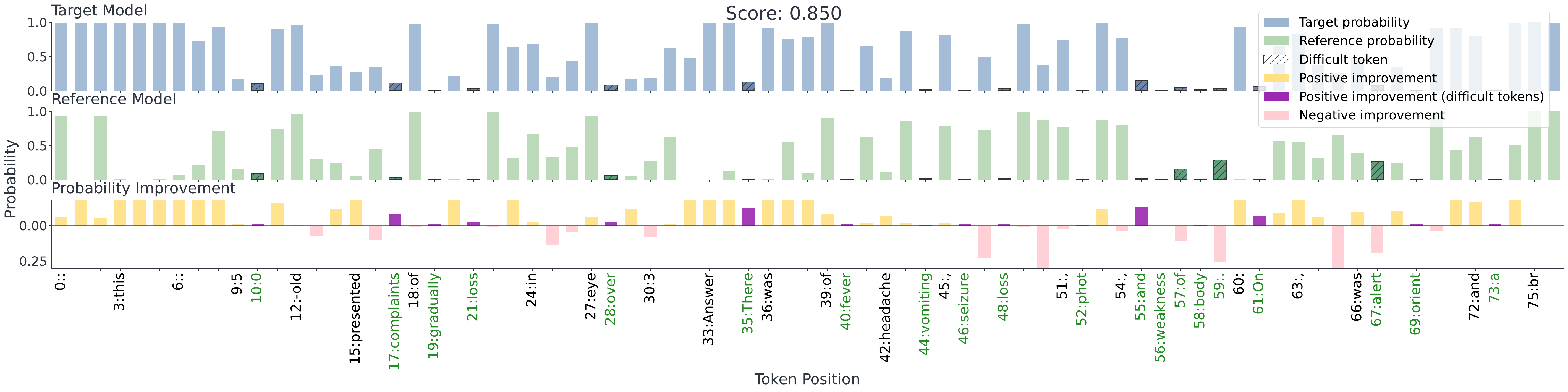}
    \caption{Non-member sample}
    \label{fig:nonmember_sample}
  \end{subfigure}

  \caption{Token-level probability improvement across target and reference model (Clinicalnotes).}
  \label{fig:token_prob_improve_cln}
\end{figure*}

\begin{figure*}[!ht]
  \centering
  \footnotesize
  \begin{subfigure}{1\textwidth}
    \includegraphics[width=\linewidth]{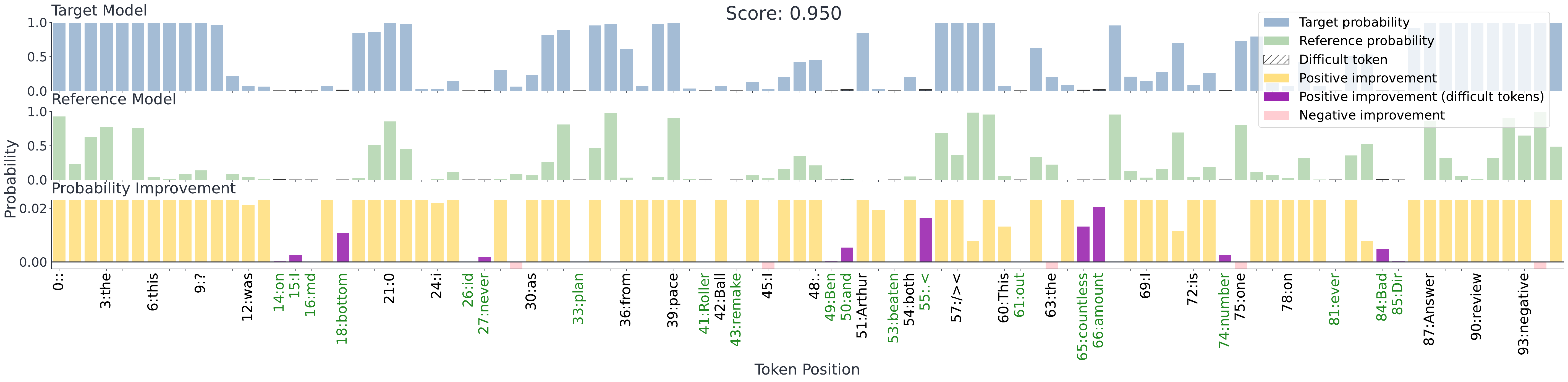}
    \caption{Member sample}
    \label{fig:member_sample_imdb}
  \end{subfigure}
  \vspace{4mm} % space between the two subfigures
  \begin{subfigure}{1\textwidth}
    \includegraphics[width=\linewidth]{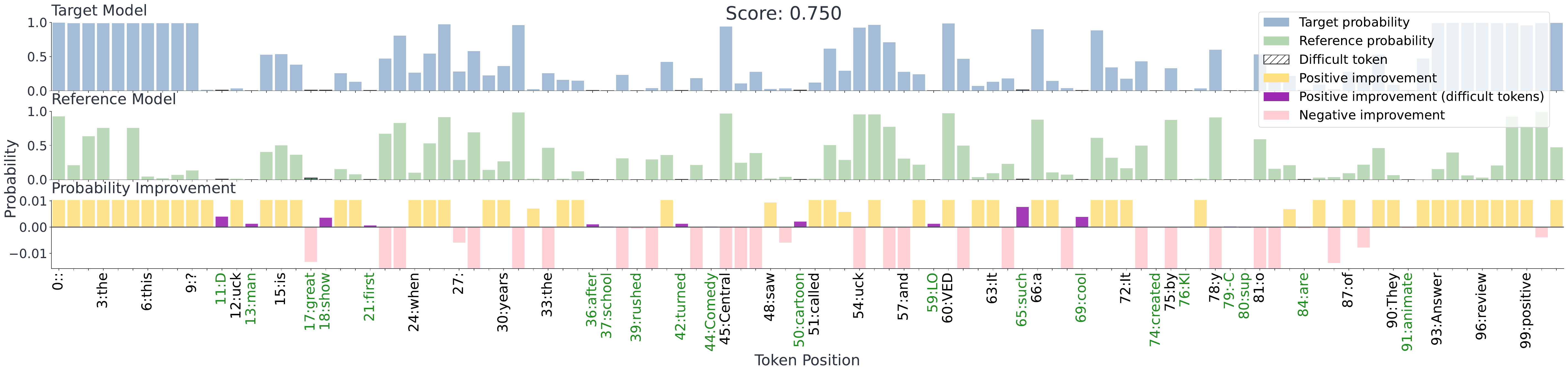}
    \caption{Non-member sample}
    \label{fig:nonmember_sample_imdb}
  \end{subfigure}
  \caption{Token-level probability improvement across target and reference model (IMDB).}
  \label{token_prob_improve_imdb}
\end{figure*}

\section{Analysis of Token-Level Probability Distribution}
\label{TokenLevelxD}
Figures~\ref{fig:token_prob_improve_cln} and~\ref{token_prob_improve_imdb} show
token-level probabilities from both target and reference models (Qwen3-0.6B) 
for member and non-member samples across Clinicalnotes and IMDB. Both samples 
show probability improvements from reference to target, even for non-members. 
However, the magnitude differs: member improvements exceed 0.2, while 
non-member improvements barely exceed 0.1. More critically, the hard-to-predict 
tokens (shown in green) reveal a pronounced distinction, roughly 91\% improve 
in member samples versus only 70\% in non-members. This fine-grained signal is 
diluted when aggregating over entire sequences, explaining why methods like 
Loss or Zlib MIA fail. By focusing on low-confidence tokens, \prjctname{} 
captures a stronger membership signal.

This pattern also explains why \prjctname{} fails to surpass Ratio on IMDB. 
IMDB non-member tokens include generic terms (e.g., \textit{man, show, school, 
cartoon}) likely seen during pretraining, whereas Clinicalnotes non-members 
contain domain-specific terms (e.g., \textit{vomiting, seizure, weakness, 
alert}). Domain-specific vocabulary makes probability improvements more 
distinct, giving \prjctname{} a clearer advantage over simpler baselines like 
Ratio.

% \section{Effect of Decision Threshold on Classification Accuracy}
% \label{sec:appendix_D}
% In Table~\ref{tab:AUC_thresholds_qwen_clinicalnotes} and~\ref{tab:AUC_thresholds_qwen_wiki} we provide a direct analysis of varying threshold by explicitly changing \(\tau\).
% As observed, very low thresholds ($\tau \in [0.1, 0.3]$) result in poor AUC performance. As the threshold increases, the AUC improves, reaching its peak at \(\tau = 0.6\) with an AUC of 80.77\%. Beyond this point, performance begins to degrade. At \(\tau = 0.7\), the AUC drops to 76.35\%, and at \(\tau = 0.9\), it decreases further to 53.77\%.

% A similar trend is observed on the Wikipedia dataset. Very low thresholds ($\tau \in [0.1, 0.2]$) again yield poor performance, with the AUC improving steadily as the threshold increases. The optimal performance is achieved at \(\tau = 0.5\) with an AUC of 83.81\%, remaining nearly stable at \(\tau = 0.6\) (83.59\%). Performance then degrades at higher thresholds, dropping to 74.68\% at \(\tau = 0.7\) and 51.58\% at \(\tau = 0.9\).

\section{Effect of Decision Threshold on Classification Accuracy}
\label{sec:appendix_D}
In Table~\ref{tab:AUC_thresholds_qwen_clinicalnotes} and~\ref{tab:AUC_thresholds_qwen_wiki} we provide a direct analysis of varying threshold by explicitly changing $\tau$. Because the evaluation set is balanced (equal numbers of members and non-members), accuracy and balanced accuracy coincide.
As observed, very low thresholds ($\tau \in [0.1, 0.3]$) result in poor accuracy. As the threshold increases, the performance improves, reaching its peak at $\tau = 0.6$ with an accuracy of $80.77\%$. Beyond this point, performance begins to degrade. At $\tau = 0.7$, the accuracy drops to $76.35\%$, and at $\tau = 0.9$, it decreases further to $53.77\%$.
A similar trend is observed on the Wikipedia dataset. Very low thresholds ($\tau \in [0.1, 0.2]$) again yield poor performance, with the accuracy improving steadily as the threshold increases. The optimal performance is achieved at $\tau = 0.5$ with an accuracy of $83.81\%$, remaining nearly stable at $\tau = 0.6$ ($83.59\%$). Performance then degrades at higher thresholds, dropping to $74.68\%$ at $\tau = 0.7$ and $51.58\%$ at $\tau = 0.9$.

\begin{table}[H]
\centering
\setlength{\tabcolsep}{4pt}
\resizebox{0.48\textwidth}{!}{%
\begin{tabular}{c c c c c c c c c c}
    \toprule
    \textbf{Threshold} & 0.1 & 0.2 & 0.3 & 0.4 & 0.5 & \textbf{0.6} & 0.7 & 0.8 & 0.9 \\
    \midrule
    \textbf{ACC} & 
    0.5015 & 
    0.5143 & 
    0.5615 & 
    0.6606 & 
    0.7676 & 
    \textbf{0.8077} & 
    0.7635 & 
    0.6425 & 
    0.5377 \\
    \bottomrule
\end{tabular}}
\caption{Classification accuracy across different decision thresholds for \prjctname{} on Qwen-3-0.6B with \textbf{Clinicalnotes} dataset.}
\label{tab:AUC_thresholds_qwen_clinicalnotes}
\end{table}

\begin{table}[H]
\centering
\setlength{\tabcolsep}{4pt}
\resizebox{0.48\textwidth}{!}{%
\begin{tabular}{c c c c c c c c c c}
    \toprule
    \textbf{Threshold} & 0.1 & 0.2 & 0.3 & 0.4 & 0.5 & \textbf{0.6} & 0.7 & 0.8 & 0.9 \\
    \midrule
    \textbf{ACC} & 
    0.5002 & 
    0.5107 & 
    0.5834 & 
    0.7286 & 
    \textbf{0.8381} & 
    0.8359 & 
    0.7468 & 
    0.6085 & 
    0.5158 \\
    \bottomrule
\end{tabular}}
\caption{Classification accuracy across different decision thresholds for \prjctname{} on Qwen-3-0.6B with \textbf{Wikipedia} dataset.}
\label{tab:AUC_thresholds_qwen_wiki}
\end{table}

% \section{Ablation Study on Hyperparameters ($\alpha$, $k_{min}$, $k_{max}$)}
\section{Ablation Study on Min-\texorpdfstring{$K$}{K} Token Selection Parameters}
\label{subsec:abltn}
Our default setting is $\alpha = 0.2$ (selecting the 20\% lowest-probability tokens), with $k_{min} = 20$ and $k_{max} = 30$, which yields a stable attack performance. These values are fixed across all datasets and models rather than tuned per benchmark, and the ablation below reports their effect on performance.

We conducted an ablation study on GPT-2 using the Clinicalnotes dataset by varying $k_{min}$, $k_{max}$, and $\alpha$. The results are presented in Table~\ref{tab:ablation_alpha}.

To isolate the effect of $\alpha$, we repeat the sweep with $k_{min} = 1$ and $k_{max}$ left uncapped, so that $k = \lfloor \alpha L \rfloor$ and $\alpha$ is the only quantity that varies across rows. Table~\ref{tab:ablation_alpha_isolated} reports the result. Performance peaks at $\alpha = 0.4$ and then declines as the selection grows toward the full sequence, which indicates that restricting the aggregation to hard tokens is what carries the membership signal. The full-sequence row removes the cap entirely and realizes the $\alpha \to 1$ limit that a finite $k_{max}$ otherwise makes unreachable. For comparison, the Loss attack reaches $0.6238$ AUC in this setting (Table~\ref{tab:auc_clinicalnotes}), below the $0.6853$ of the full-sequence row of Table~\ref{tab:ablation_alpha_isolated}, showing that sign aggregation is not equivalent to loss aggregation.

\begin{table}[H]
\centering
\small
\resizebox{\columnwidth}{!}{%
\begin{tabular}{cccccc}
\toprule
$k_{max}$ & $k_{min}$ & $100\alpha$ (\%) & AUC & TPR@FPR=0.1 & TPR@FPR=0.01 \\
\midrule
5   & 2  & 1--5 & 0.6312--0.6588 & 0.27--0.34 & 0.0000 \\
5   & 2  & 20   & 0.6631         & 0.3477     & 0.0005 \\
20  & 10 & 20   & 0.6946         & 0.3098     & 0.0568 \\
30  & 20 & 20   & 0.6978         & 0.3223     & 0.0524 \\
80  & 70 & 20   & 0.6771         & 0.2614     & 0.0426 \\
100 & 90 & 20   & 0.6824         & 0.2794     & 0.0350 \\
100 & 90 & 100  & 0.6854         & 0.2840     & 0.0385 \\
\bottomrule
\end{tabular}}
\caption{Ablation study results on GPT-2 (Clinicalnotes) across varying $k_{max}$, $k_{min}$, and $\alpha$. The first row summarises five individual settings, $100\alpha = 1$ to $5$, and reports the range spanned by them rather than a single configuration.}
\label{tab:ablation_alpha}
\end{table}

\begin{table}[H]
\centering
\small
\resizebox{0.98\columnwidth}{!}{%
\begin{tabular}{c c c c}
\toprule
$100\alpha$ (\%) & AUC & TPR@FPR=0.1 & TPR@FPR=0.01 \\
\midrule
5   & 0.6629 & 0.2504 & 0.0000 \\
10  & 0.6898 & 0.3013 & 0.0698 \\
20  & 0.6924 & 0.2975 & 0.0672 \\
40  & \textbf{0.7030} & 0.3176 & 0.0596 \\
60  & 0.6947 & 0.3052 & 0.0438 \\
80  & 0.6883 & 0.2945 & 0.0388 \\
\midrule
100 & 0.6853 & 0.2786 & 0.0271 \\
\bottomrule
\end{tabular}}
\caption{Effect of $\alpha$ in isolation on \prjctname{} (GPT-2, Clinicalnotes). Unlike Table~\ref{tab:ablation_alpha}, here we fix $k_{min}=1$ and leave $k_{max}$ uncapped, so that $k=\lfloor \alpha L \rfloor$ and $\alpha$ is the only quantity that varies. The last row ($\alpha=1$) selects every token, realizing the full-sequence limit that a finite $k_{max}$ otherwise makes unreachable.}
\label{tab:ablation_alpha_isolated}
\end{table}

\section{Probability versus Log-Probability Space}
\label{sec:appendix_problogprob}
\prjctname{} aggregates token probabilities only through their ordering (for token selection) and the sign of their difference (for scoring). Both are preserved under the logarithm, since $\log p_{\F,i} > \log p_{\R,i}$ exactly when $p_{\F,i} > p_{\R,i}$. The attack is therefore invariant to whether we compare in probability space, $\Delta_i = p_{\F,i} - p_{\R,i}$, or log-probability space, $\Delta^{\log}_i = \log p_{\F,i} - \log p_{\R,i}$. Table~\ref{tab:prob_vs_logprob} confirms this: the two variants agree to within $0.0002$ AUC, well below the bootstrap standard deviation ($\pm 0.0020$). We use probability space in all experiments.

\begin{table}[H]
\centering
\small
\setlength{\tabcolsep}{6pt}
\resizebox{0.48\textwidth}{!}{%
\begin{tabular}{l | c c c}
    \toprule
    \textbf{Aggregation space} & \textbf{AUC} & \textbf{TPR@FPR=0.1} & \textbf{TPR@FPR=0.01} \\
    \midrule
    Probability: $\Delta_i = p_{\F,i} - p_{\R,i}$ & 0.8843 & 0.6861 & 0.3839 \\
    Log-probability: $\Delta_i = \log p_{\F,i} - \log p_{\R,i}$ & 0.8845 & 0.6851 & 0.3810 \\
    \bottomrule
\end{tabular}}
\caption{Probability- versus log-probability-space aggregation for \prjctname{}, with Qwen-3-0.6B fine-tuned on Clinicalnotes as the target and Qwen-3-0.6B-Base as the reference. The two variants are equivalent by construction; the small differences come from floating-point rounding when $p_{\F,i}$ and $p_{\R,i}$ are nearly equal.}
\label{tab:prob_vs_logprob}
\end{table}

% \begin{table*}[t]
% \centering
% \caption{Ablation study results on GPT-2 (ClinicalNotes) across varying $k_{max}$, $k_{min}$, and $\alpha$.}
% \label{tab:ablation_alpha}
% \begin{tabular}{cccccc}
% \toprule
% $k_{max}$ & $k_{min}$ & $\alpha$ (\%) & AUC & TPR@FPR=0.1 & TPR@FPR=0.01 \\
% \midrule
% 5   & 2  & 1--5 & 0.6312--0.6588 & 0.27--0.34 & 0.0000 \\
% 5   & 2  & 20   & 0.6631         & 0.3477     & 0.0005 \\
% 20  & 10 & 20   & 0.6946         & 0.3098     & 0.0568 \\
% 30  & 20 & 20   & 0.6978         & 0.3223     & 0.0524 \\
% 80  & 70 & 20   & 0.6771         & 0.2614     & 0.0426 \\
% 100 & 90 & 20   & 0.6824         & 0.2794     & 0.0350 \\
% 100 & 90 & 100  & 0.6854         & 0.2840     & 0.0385 \\
% \bottomrule
% \end{tabular}
% \end{table*}

Overall, \prjctname{} is stable across a broad range of hyperparameters. With $\alpha = 0.2$, AUC stays consistent across many choices of $k_{min}$ and $k_{max}$, and the default setting $(20, 30)$ lies in this stable region. Performance degrades only when $K$ becomes very small (notably at 
$K=2$ and $K=5$), where TPR@FPR=0.01 collapses due to too few tokens. Increasing $K$ beyond the default range gradually reduces AUC, indicating that adding higher-probability tokens dilutes the signal. Reducing $\alpha$ below $0.2$ also weakens performance. Because a finite $k_{max}$ caps the selection at $k_{max}$ tokens, raising $\alpha$ in this table does not reach the full-sequence limit; Table~\ref{tab:ablation_alpha_isolated} removes the cap and shows that extending the selection to all tokens ($\alpha = 1$) reduces performance. These trends confirm that $\alpha = 0.2$ and focusing on low-probability tokens is a robust default choice.

\section{Effect of Model Choice on Token Selection}
\label{sec:appendix_E}
In our main experiment, the \prjctname{} algorithm uses the target model’s token-level probability distribution to identify the most problematic tokens. In Table~\ref{tab:ablation_token_selection} we experiment by switching the algorithm to use the reference model for token selection.
\begin{table}[H]
\centering
\setlength{\tabcolsep}{6pt}
\begin{subtable}[t]{0.42\textwidth}
\centering
\resizebox{\textwidth}{!}{%
\begin{tabular}{l | c c}
    \toprule
    \textbf{Token Selection Strategy} & \textbf{Qwen-3-0.6B} & \textbf{LLaMA-3.2-1B} \\
    \midrule
    Target Model     & 0.8843 & 0.8659 \\
    Reference Model  & 0.8648 & 0.8474 \\
    \bottomrule
\end{tabular}}
\caption{AUC}
\label{tab:ablation_auc}
\end{subtable}
\hfill
\begin{subtable}[t]{0.42\textwidth}
\centering
\resizebox{\textwidth}{!}{%
\begin{tabular}{l | c c}
    \toprule
    \textbf{Token Selection Strategy} & \textbf{Qwen-3-0.6B} & \textbf{LLaMA-3.2-1B} \\
    \midrule
    Target Model     & 0.6861 & 0.7057 \\
    Reference Model  & 0.6999 & 0.6901 \\
    \bottomrule
\end{tabular}}
\caption{TPR@FPR=0.1}
\label{tab:ablation_tpr01}
\end{subtable}
\hfill
\begin{subtable}[t]{0.42\textwidth}
\centering
\resizebox{\textwidth}{!}{%
\begin{tabular}{l | c c}
    \toprule
    \textbf{Token Selection Strategy} & \textbf{Qwen-3-0.6B} & \textbf{LLaMA-3.2-1B} \\
    \midrule
    Target Model     & 0.3839 & 0.4253 \\
    Reference Model  & 0.2986 & 0.3545 \\
    \bottomrule
\end{tabular}}
\caption{TPR@FPR=0.01}
\label{tab:ablation_tpr001}
\end{subtable}

\caption{Effect of the target model vs. the reference model on token selection on Target models fine-tuned on Clinicalnotes dataset.}
\label{tab:ablation_token_selection}
\end{table}

Both in terms of AUC and TPR@FPR=0.01, the performance shows a slight decrease (AUC drops by about 2.1\% for LLaMA-3.2-1B). However, for the more relaxed case of TPR@FPR=0.1, Qwen exhibits a small improvement (about 2.0\%), whereas LLaMA-3 still decreases slightly (around 2.2\%).

\section{Effect of Reference Model on \prjctname{} Performance}
\label{RefModelxC}
For our primary results, we use the same base pretrained version of the target model as the reference; in practice an attacker may not have this exact model. We therefore test two alternatives (Tables~\ref{tab:ref_model_pim_mia} and~\ref{tab:ref_model_pim_mia_gpt2}), in both cases encoding the input with the target's tokenizer and scoring the same token-ID sequence with both models. A matched-tokenizer reference from the same family (Qwen-2.5-0.5B for Qwen-3-0.6B, DistilGPT-2 for GPT-2) shares the target's vocabulary, and performance drops only moderately: the AUC decreases from 88.43\% to 87.20\% and from 69.77\% to 66.74\%, respectively. A mismatched-tokenizer reference from a different family (Gemma-2-2B and Qwen-3-0.6B, denoted $^{\dagger}$) has a different vocabulary, so token-level probabilities cannot be aligned and \prjctname{} drops sharply (AUC 52.50\% and 60.00\%). A shared or exactly alignable tokenizer is therefore a requirement of the method.
% \begin{table}[!ht]
% \centering
% \setlength{\tabcolsep}{6pt}
% \resizebox{0.48\textwidth}{!}{%
% \begin{tabular}{l c c c}
%     \toprule
%     \textbf{Reference Model} & \textbf{AUC} & \textbf{TPR@FPR=0.1} & \textbf{TPR@FPR=0.01} \\
%     \midrule
%     Qwen3-0.6B        & $0.8843 \pm 0.0019$ & $0.6861 \pm 0.0046$ & $0.3839 \pm 0.0056$ \\
%     Qwen2.5-0.5B      & $0.8720 \pm 0.0021$ & $0.6818 \pm 0.0044$ & $0.2765 \pm 0.0083$ \\
%     \bottomrule
% \end{tabular}}
% % \caption{Effect of Reference model on PIM-MIA performance on Target model \texttt{Qwen3-0.6B} fine-tuned on \texttt{Asclepius} dataset.}
% \caption{Effect of Reference model on PIM-MIA performance on Target model Qwen3-0.6B fine-tuned on Clinicalnotes dataset.}
% \label{tab:ref_model_pim_mia}
% \end{table}

% \begin{table}[!ht]
% \centering
% \setlength{\tabcolsep}{6pt}
% \resizebox{0.48\textwidth}{!}{%
% \begin{tabular}{l c c c}
%     \toprule
%     \textbf{Reference Model} & \textbf{AUC} & \textbf{TPR@FPR=0.1} & \textbf{TPR@FPR=0.01} \\
%     \midrule
%     GPT-2        & $0.6977 \pm 0.0030$ & $0.3222 \pm 0.0038$ & $0.0522 \pm 0.0124$ \\
%     DistilGPT-2  & $0.6674 \pm 0.0031$ & $0.2603 \pm 0.0068$ & $0.0540 \pm 0.0032$ \\
%     \bottomrule
% \end{tabular}}
% \caption{Effect of Reference model on PIM-MIA performance on Target model GPT-2 fine-tuned on Clinicalnotes dataset.}
% \label{tab:ref_model_pim_mia_gpt2}
% \end{table}

\begin{table}[H]
\centering
\setlength{\tabcolsep}{6pt}
\resizebox{0.48\textwidth}{!}{%
\begin{tabular}{l | c c c}
    \toprule
    \textbf{Reference Model} & \textbf{AUC} & \textbf{TPR@FPR=0.1} & \textbf{TPR@FPR=0.01} \\
    \midrule
    Qwen-3-0.6B        & 0.8843 & 0.6861 & 0.3839 \\
    Qwen-2.5-0.5B      & 0.8720 & 0.6818 & 0.2765 \\
    Gemma-2-2B$^{\dagger}$  & 0.5250 & 0.0000 & 0.0000 \\
    \bottomrule
\end{tabular}}
\caption{Effect of Reference model on \prjctname{} performance on Target model Qwen-3-0.6B fine-tuned on Clinicalnotes dataset. $^{\dagger}$~denotes a reference model with a mismatched tokenizer.}
\label{tab:ref_model_pim_mia}
\end{table}

\begin{table}[H]
\centering
\setlength{\tabcolsep}{6pt}
\resizebox{0.48\textwidth}{!}{%
\begin{tabular}{l | c c c}
    \toprule
    \textbf{Reference Model} & \textbf{AUC} & \textbf{TPR@FPR=0.1} & \textbf{TPR@FPR=0.01} \\
    \midrule
    GPT-2        & 0.6978 & 0.3223 & 0.0524 \\
    DistilGPT-2  & 0.6674 & 0.2603 & 0.0540 \\
    Qwen-3-0.6B$^{\dagger}$ & 0.6000 & 0.0000 & 0.0000 \\
    \bottomrule
\end{tabular}}
\caption{Effect of Reference model on \prjctname{} performance on Target model GPT-2 fine-tuned on Clinicalnotes dataset. $^{\dagger}$~denotes a reference model with a mismatched tokenizer.}
\label{tab:ref_model_pim_mia_gpt2}
\end{table}

% \section{Effecnt on post fine-tuning performance due to dp-sgd}
% \label{sec:appendix_F}
% \begin{table}[t]
% \centering
% \small
% \resizebox{0.48\textwidth}{!}{%
% \begin{tabular}{lcccccc}
% \toprule
% & \multicolumn{3}{c}{\textbf{Qwen-3-0.6B}} & \multicolumn{3}{c}{\textbf{GPT-2}} \\
% \cmidrule(lr){2-4} \cmidrule(lr){5-7}
% \textbf{Task} & \textbf{Base} & \textbf{FT} & \textbf{DP} & \textbf{Base} & \textbf{FT} & \textbf{DP} \\
% \midrule
% MedMCQA           & 0.4224 & 0.2890 & 0.2893 & 0.2426 & 0.2921 & 0.2649 \\
% Clinical Knowledge & 0.5736 & 0.2679 & 0.2151 & 0.2302 & 0.2642 & 0.2453 \\
% College Biology    & 0.6181 & 0.2569 & 0.2569 & 0.2569 & 0.2929 & 0.2361 \\
% PubMedQA          & 0.6580 & 0.5580 & 0.4620 & 0.4800 & 0.4800 & 0.3240 \\
% \bottomrule
% \end{tabular}
% }
% \caption{Three-shot downstream task accuracy comparison between Qwen-3-0.6B and GPT-2 across medical QA benchmarks. FT denotes fine-tuned models, and DP denotes DP fine-tuned models, both trained on the Asclepius dataset.}
% \label{tab:combined_3shot_asclepius}
% \end{table}

\section{Robustness Across Training Seeds}\label{sec:appendix_seed}

To test how sensitive our results are to the training seed, we fine-tune
Qwen-3-0.6B on Clinicalnotes with two additional seeds ($7$ and $13$) and
repeat the full attack evaluation. The main paper uses seed $42$, the default
in our training script. Every other factor is held fixed across the three
runs: the member and non-member split, the training hyperparameters, the
evaluation set, and the reference model. We evaluate five attacks and report
AUC as the bootstrap mean over $1000$ resamples of the evaluation set.

\begin{table}[!ht]
\centering
\small
\setlength{\tabcolsep}{4pt}
\resizebox{\columnwidth}{!}{%
\begin{tabular}{l c c c}
\toprule
\textbf{Attack} & \textbf{Seed 42} & \textbf{Seed 7} & \textbf{Seed 13} \\
\midrule
\prjctname{}   & \textbf{0.8843 $\pm$ 0.0019} & \textbf{0.8279 $\pm$ 0.0024} & \textbf{0.8272 $\pm$ 0.0023} \\
Ratio          & 0.8681 $\pm$ 0.0021          & 0.8069 $\pm$ 0.0026          & 0.8087 $\pm$ 0.0025 \\
Loss           & 0.8100 $\pm$ 0.0024          & 0.7406 $\pm$ 0.0027          & 0.7416 $\pm$ 0.0028 \\
Zlib           & 0.7838 $\pm$ 0.0026          & 0.7148 $\pm$ 0.0030          & 0.7160 $\pm$ 0.0030 \\
Lowercase      & 0.7332 $\pm$ 0.0027          & 0.6617 $\pm$ 0.0032          & 0.6630 $\pm$ 0.0031 \\
\bottomrule
\end{tabular}}
\caption{AUC across three training seeds for Qwen-3-0.6B fine-tuned on
Clinicalnotes. Seed $42$ is our default seed; seeds $7$ and $13$ are
independent reruns under the same recipe. Values are bootstrap means $\pm$ one
standard deviation.}
\label{tab:seed_study}
\end{table}

Table~\ref{tab:seed_study} shows that the ranking of the five attacks is
identical under all three seeds: \prjctname{} ranks first, followed by Ratio,
Loss, Zlib, and Lowercase. Absolute AUC is lower under seeds $7$ and $13$ than
under seed $42$ by a near-uniform $0.057$ to $0.071$ across all attacks. The
comparison between methods is unaffected: \prjctname{} preserves its margin
over the strongest baseline under both additional seeds. Absolute attack success thus depends
on the specific fine-tuned checkpoint, while the relative comparison is stable
across seeds.

\section{Analysis of the Differential Privacy Defense}
\label{sec:appendix_F}

% \begin{table}[!ht]
% \centering
% \small
% \resizebox{0.48\textwidth}{!}{%
% \begin{tabular}{l | cc | cc}
% \toprule
% & \multicolumn{2}{c}{\textbf{Qwen-3-0.6B}} & \multicolumn{2}{c}{\textbf{GPT-2}} \\
% \cmidrule(lr){2-3} \cmidrule(lr){4-5}
% \textbf{Task} & \textbf{FT} & \textbf{DP} & \textbf{FT} & \textbf{DP} \\
% \midrule
% MedMCQA           & 0.2890 & 0.2893 & 0.2921 & 0.2649 \\
% Clinical Knowledge & 0.2679 & 0.2151 & 0.2642 & 0.2453 \\
% College Biology    & 0.2569 & 0.2569 & 0.2929 & 0.2361 \\
% PubMedQA          & 0.5580 & 0.4620 & 0.4800 & 0.3240 \\
% \bottomrule
% \end{tabular}
% }
% \caption{Three-shot downstream task accuracy comparison between standard fine-tuning (FT) and DP fine-tuning on the Asclepius dataset for Qwen-3-0.6B and GPT-2.}
% \label{tab:ft_dpsgd_comparison}
% \end{table}

To evaluate the privacy-utility tradeoff, we measure accuracy on four downstream medical QA tasks (MedMCQA, Clinical Knowledge, College Biology, and PubMedQA~\cite{pal2022medmcqa, hendrycks2020measuring, jin2019pubmedqa}), each in a 3-shot setting with the same prompt format and evaluation split for every model. As shown in Table~\ref{tab:ft_dpsgd_comparison}, DP fine-tuning reduces utility for both models. For GPT-2, accuracy drops on most tasks, by 9.3\% on MedMCQA, 7.2\% on Clinical Knowledge, 19.4\% on College Biology, and 32.5\% on PubMedQA. For Qwen-3-0.6B, DP has little effect on MedMCQA and College Biology but degrades Clinical Knowledge (19.7\%) and PubMedQA (17.2\%). The base-model columns add context: for GPT-2, fine-tuning improves three of the four tasks, so the DP drop is a cost on top of a real gain; for Qwen-3-0.6B, the base model outperforms both fine-tuned variants on every task (e.g., $57.4\%$ vs.\ $26.8\%$ on Clinical Knowledge), so most of the utility loss comes from adapting a strong base model to the medical corpus, with DP adding to it rather than being the main cause.

\begin{table}[!ht]
\centering
\small
\resizebox{0.48\textwidth}{!}{%
\begin{tabular}{l | ccc | ccc}
\toprule
& \multicolumn{3}{c}{\textbf{Qwen-3-0.6B}} & \multicolumn{3}{c}{\textbf{GPT-2}} \\
\cmidrule(lr){2-4} \cmidrule(lr){5-7}
\textbf{Task} & \textbf{Base} & \textbf{FT} & \textbf{DP} & \textbf{Base} & \textbf{FT} & \textbf{DP} \\
\midrule
MedMCQA            & 0.4215 & 0.2890 & 0.2893 & 0.2429 & 0.2921 & 0.2649 \\
Clinical Knowledge & 0.5736 & 0.2679 & 0.2151 & 0.2302 & 0.2642 & 0.2453 \\
College Biology    & 0.6181 & 0.2569 & 0.2569 & 0.2569 & 0.2929 & 0.2361 \\
PubMedQA           & 0.6580 & 0.5580 & 0.4620 & 0.4820 & 0.4800 & 0.3240 \\
\bottomrule
\end{tabular}
}
\caption{Three-shot downstream task accuracy for the untuned base model (Base), standard fine-tuning (FT), and DP fine-tuning (DP) on the Asclepius dataset, for Qwen-3-0.6B and GPT-2. DP denotes the preliminary $\varepsilon=100$, $\delta=10^{-5}$, $C=100$ intervention of Figure~\ref{fig:radar_dpsgd} rather than a formally accounted operating point; the $C=1$ sweep of Table~\ref{tab:dp_sweep} is a separate experiment.}
\label{tab:ft_dpsgd_comparison}
\end{table}

To test whether the defense holds beyond a single setting, we fine-tune GPT-2 on Clinicalnotes with DP at four \emph{target} budgets, $\varepsilon \in \{1,2,4,8\}$, with clipping norm $C=1$ and $\delta=10^{-5}$ (Table~\ref{tab:dp_sweep}). The targets are design parameters used only to calibrate the noise multiplier; the guarantees we report throughout are the \emph{achieved} budgets returned by the accountant at the end of training, $\varepsilon \in \{0.22, 0.79, 2.84, 6.37\}$, and all four runs therefore satisfy a budget at least as tight as their target. The $\varepsilon=\infty$ row is a non-DP control trained under the same three-epoch schedule; it applies neither DP noise nor per-example clipping, so the comparison against it reflects the combined effect of the DP mechanism rather than of noise alone. Across the four DP rows it is the noise multiplier that varies, giving each row its different achieved budget. This control is highly vulnerable (\prjctname{} AUC $0.82$), while every DP budget reduces all attacks to near chance ($0.50$ to $0.51$); the reduction is already saturated at the loosest achieved budget ($\varepsilon=6.37$), so DP reduces attack success across the tested range rather than at a single operating point. Because this three-epoch schedule differs from the single-epoch recipe of our main experiments, these absolute values ($0.82$ vs.\ $0.6978$ in Table~\ref{tab:allmetrics_clinicalnotes}) are comparable within the sweep but not across the two settings. Table~\ref{tab:dp_accounting} reports the full privacy accounting for each budget. All runs use the Opacus RDP accountant with Poisson subsampling (add-or-remove-one adjacency), clipping norm $C=1$, $\delta=10^{-5}$, $N=15{,}000$ training examples, logical and physical batch size $4$ (sampling rate $q = 2.67 \times 10^{-4}$), and $3$ epochs ($11{,}250$ steps).

\begin{table}[!ht]
\centering
\small
\resizebox{0.48\textwidth}{!}{%
\begin{tabular}{c c | c c c}
\toprule
Achieved $\varepsilon$ & Target $\varepsilon$ & \prjctname{} & Loss & Ratio \\
\midrule
$\infty$ (non-DP) & --- & 0.8204 & 0.7680 & 0.7936 \\
\midrule
6.37 & 8 & 0.5087 & 0.5036 & 0.4919 \\
2.84 & 4 & 0.5020 & 0.5041 & 0.4918 \\
0.79 & 2 & 0.5093 & 0.5054 & 0.4941 \\
0.22 & 1 & 0.5020 & 0.5056 & 0.4919 \\
\bottomrule
\end{tabular}}
\caption{MIA AUC on GPT-2 fine-tuned on Clinicalnotes with DP ($C=1$, $\delta=10^{-5}$) at four privacy budgets. Rows are labelled by the achieved $\varepsilon$ reported by the accountant, with the target $\varepsilon$ used for calibration given alongside; $\varepsilon=\infty$ is the non-DP control.}
\label{tab:dp_sweep}
\end{table}

\begin{table}[!ht]
\centering
\small
\resizebox{0.48\textwidth}{!}{%
\begin{tabular}{c c c}
\toprule
Achieved $\varepsilon$ & Target $\varepsilon$ & Noise multiplier $\sigma$ \\
\midrule
0.22 & 1 & 0.781 \\
0.79 & 2 & 0.607 \\
2.84 & 4 & 0.485 \\
6.37 & 8 & 0.398 \\
\bottomrule
\end{tabular}}
\caption{DP accounting for the privacy sweep of Table~\ref{tab:dp_sweep}. Achieved $\varepsilon$ is the budget the accountant reports at the end of training and is the guarantee we report; target $\varepsilon$ is the design budget from which the noise multiplier $\sigma$ was calibrated before training. Achieved $\varepsilon$ falls below its target in every run, so each guarantee is conservative with respect to the budget it was designed for.}
\label{tab:dp_accounting}
\end{table}

A natural question is whether DP suppresses learning entirely or leaves the model able to fit the domain without memorizing it. Table~\ref{tab:dp_contamination} reports PPL on training members and held-out non-members for the base model, standard fine-tuning, and each DP model. The base model assigns similar high PPL to both splits ($51.9$ and $52.3$). Standard fine-tuning lowers member PPL to $7.7$ and opens a member/non-member gap ($7.7$ vs.\ $11.9$). Under DP, PPL falls to between $27$ and $33$ depending on the achieved budget, well below the base model, while the member/non-member gap closes to within $0.1$, and PPL rises as the achieved $\varepsilon$ decreases (from $26.6$ at $\varepsilon=6.37$ to $33.1$ at $\varepsilon=0.22$). In these measurements, DP lowers PPL relative to the base model while the member/non-member gap nearly vanishes. This is a targeted check on one model and one dataset (GPT-2 on Clinicalnotes). Comparing against the base checkpoint controls for the probability changes fine-tuning induces. It does not show that evaluation examples were absent from pretraining. We do not quantify or rule out such overlap.

\begin{table}[!ht]
\centering
\small
\resizebox{0.48\textwidth}{!}{%
\begin{tabular}{l c | c c}
\toprule
Model & Target $\varepsilon$ & PPL (member) & PPL (non-member) \\
\midrule
Base GPT-2 & --- & 51.9 & 52.3 \\
Standard FT ($\varepsilon=\infty$) & --- & 7.7 & 11.9 \\
\midrule
DP $\varepsilon=6.37$ & 8 & 26.6 & 26.7 \\
DP $\varepsilon=2.84$ & 4 & 28.3 & 28.4 \\
DP $\varepsilon=0.79$ & 2 & 30.4 & 30.5 \\
DP $\varepsilon=0.22$ & 1 & 33.1 & 33.2 \\
\bottomrule
\end{tabular}}
\caption{PPL on training members and held-out non-members (GPT-2, Clinicalnotes) for the base model, standard fine-tuning, and DP models. DP rows are labelled by the achieved $\varepsilon$, with the target $\varepsilon$ used for calibration given alongside.}
\label{tab:dp_contamination}
\end{table}

\end{document}